\documentclass[aps,prd,twocolumn,floatfix,showpacs,10pt,superscriptaddress,preprintnumbers,nofootinbib]{revtex4-1}

\usepackage{graphicx}
\usepackage{amsmath,amssymb,bm}
\usepackage{xcolor}
\usepackage{tikz}
\usetikzlibrary{arrows.meta}
\usepackage[colorlinks=true,linkcolor=blue,citecolor=blue,urlcolor=blue]{hyperref}

\graphicspath{{figs/}}
\definecolor{darkyellow}{rgb}{0.5, 0.5, 0.0}
\definecolor{darkpurple}{rgb}{0.5, 0.2, 0.8}
\definecolor{darkblue}{rgb}{0.0, 0.0, 0.8}
\definecolor{darkgreen}{rgb}{0.0, 0.4, 0.0}
\definecolor{darkred}{rgb}{0.5, 0.0, 0.0}

\definecolor{ReviewOrange}{RGB}{205,102,0}

\usepackage{marginnote}
\usepackage[normalem]{ulem}

\newcommand{\Ecal}{\mathcal{E}}
\newcommand{\Kcal}{\mathcal{K}}
\newcommand{\Wcal}{\mathcal{W}}
\newcommand{\Jcal}{\mathcal{J}}
\newcommand{\Fcal}{\mathcal{F}}
\newcommand{\dd}{\mathrm{d}}

\begin{document}

\preprint{SLAC-PUB-261006}
\preprint{MIT-CTP 6127}

\title{Precision Studies of Boosted Polarized Vector Bosons with Energy Correlators}

\author{Anjie Gao}
\affiliation{SLAC National Accelerator Laboratory, Stanford University, Stanford, CA 94309, USA}
\email{anjiegao@stanford.edu}
\author{Kyle Lee}
\affiliation{High Energy Physics Division, Argonne National Laboratory, Lemont, IL 60439, USA}
\email{kyle@anl.gov}
\author{Xiaoyuan Zhang}
\affiliation{Center for Theoretical Physics -- a Leinweber Institute, Massachusetts Institute of Technology, Cambridge, MA 02139, USA}
\email{xyz2@mit.edu}

\begin{abstract}
Boosted, hadronically decaying $W$ and $Z$ bosons are among the most abundant heavy objects at colliders, and their vector nature makes their substructure qualitatively richer than that of a scalar.
We initiate precision studies of the hadronic decays of boosted polarized vector bosons with energy correlators.
In the narrow-width limit, the laboratory-frame two-point energy correlator is a sum of longitudinal and transverse decay spectra, each weighted by its polarized production rate and obtained by a closed-form boost kernel acting on rest-frame correlators. Combining polarized production density matrices, computed at NNLO in QCD for $pp\to V+X$ and at leading order for $e^+e^-\to W^+W^-$, with decay correlators that combine NNLO fixed order with NNLL collinear and N$^4$LL back-to-back resummation, we find for high-$p_T$ $W/Z$ production that polarization shifts the spectrum by $\mathcal{O}(10\%)$ relative to a naive unpolarized treatment.
Even without any reconstruction, hadronic $W/Z$ decays stand out over the realistic QCD background of untagged jet samples as a localized angular bump. Our results open the door to using energy correlators for polarimetry of electroweak bosons through their hadronic decays, for standard-candle measurements of electroweak bosons inside jets, and in sharper searches for new resonances.
\end{abstract}

\maketitle
\makeatletter\def\l@f@section{\addpenalty{\@secpenalty}\addvspace{0.5em plus\p@}}\makeatother
\tableofcontents

\section{Introduction}
\label{sec:intro}

At the LHC, electroweak (EW) bosons are routinely produced with energies far above their masses, collimating their dominant hadronic decays into single jets.
Resolving the internal structure of such jets has grown into the mature field of jet substructure~\cite{Butterworth:2008iy,Larkoski:2017jix}.
Within it, energy correlators, the correlation functions $\langle \Ecal(\vec n_1)\Ecal(\vec n_2)\cdots\rangle$ of the asymptotic energy flow operator~\cite{Sterman:1975xv, Sveshnikov:1995vi, Tkachov:1995kk, Korchemsky:1999kt, Hofman:2008ar,Moult:2025nhu}
\begin{align}\label{eq:Edef}
\Ecal(\vec n) = \lim_{r\to\infty}\int_0^{\infty} \dd t\; r^2\, n^i\, T_{0i}(t, r\vec n)\,,
\end{align}
have emerged as jet substructure observables with a direct field-theoretic definition, enabling precision calculations and precision measurements inside QCD jets at hadron colliders~\cite{Lee:2022uwt,Chen:2023zlx,CMS:2024mlf,ALICE:2024dfl,STAR:2025jut,Lee:2026hub} and in the collinear limit at lepton colliders~\cite{Dixon:2019uzg,Chen:2020vvp,Jaarsma:2023ell,Jaarsma:2025tck,Electron-PositronAlliance:2025fhk,Lee:2026zyl}.

Very recently, the precision study of \emph{boosted heavy-particle decays} with energy correlators was initiated for the Higgs boson~\cite{Gao:2026xuq} and the unpolarized $Z$ boson~\cite{Holguin:2026vld}.
There, it was shown that the two-point energy--energy correlator (EEC) of a boosted decaying resonance, measured as a function of the lab-frame detector opening angle, exhibits a characteristic decay peak at
\begin{align}
z' \simeq \frac{1}{\gamma^2}\,, \qquad z' \equiv \frac{1-\cos\theta'}{2}\,,
\end{align}
with $\gamma = E_V/m_V$ the boost. Here and below, primes denote laboratory-frame quantities and unprimed symbols their rest-frame counterparts. The decay peak imprints the resonance mass on the angular distribution of energy flux and breaks the approximate scale invariance of QCD jets.\footnote{At hadron colliders, angular separations are often quantified using the longitudinally boost-invariant distance $\Delta R = \sqrt{\Delta\eta^2+\Delta\phi^2}$, which coincides with the opening angle $\theta'$ at small angles for central rapidities, so that the decay peak sits at $\Delta R \simeq 2/\gamma \simeq 2m_V/p_T$.} 

Boosted $W$ and $Z$ bosons are an ideal laboratory for this program.
They are produced at rates orders of magnitude beyond the Higgs~\cite{ATLAS:2026iwc}, in inclusive $V+$jets production, in diboson processes, and in top decays, and are routinely identified as boosted objects in experimental analyses~\cite{Kogler:2018hem,CMS:2014rsx,ATLAS:2015lxh}.
Importantly, unlike the Higgs boson, they carry \emph{spin}, and precision predictions for their substructure must account for it. For a \emph{scalar} or an unpolarized vector boson, the rest-frame decay is isotropic, and the boosted spectrum is an exact one-dimensional convolution of the rest-frame EEC with a single universal boost kernel~\cite{Gao:2026xuq}:
\begin{align}\label{eq:scalarboost}
\frac{\dd\sigma}{\dd z'} = \int_0^1 \dd z\; \Kcal_\gamma(z,z')\,\frac{\dd\sigma}{\dd z}\,,
\end{align}
where $\dd\sigma/\dd z$ is the rest-frame EEC in the analogous variable $z=(1-\cos\theta)/2$ and $\Kcal_\gamma$ is a purely kinematic kernel known in closed form.
However, an unpolarized vector boson is an idealization that no realistic production mechanism realizes, so the single convolution given in Eq.~\eqref{eq:scalarboost} no longer suffices.

A produced vector boson is characterized by a spin density matrix $\rho_{\lambda\lambda'}$ over its helicity states, whose measurement is an important target of electroweak physics at colliders.
At high energies, amplitudes involving longitudinally polarized
$W$ and $Z$ bosons are related to those involving the would-be
Goldstone bosons of electroweak symmetry breaking~\cite{Cuomo:2019siu}.
Longitudinal polarization therefore provides a probe of
electroweak symmetry breaking, while polarization fractions
and helicity-interference observables more generally probe
electroweak interactions and possible anomalous couplings~\cite{Hagiwara:1986vm,Panico:2017frx,Azatov:2016sqh}. At the LHC, polarization measurements have relied predominantly
on leptonic decays~\cite{CMS:2011kaj,ATLAS:2012au, ATLAS:2019bsc, CMS:2021icx}. In Ref.~\cite{Ricci:2022htc}, it was demonstrated at leading order that the polarization dependence is cleanly imprinted on energy correlators measured on the hadronic decays of boosted vector bosons. In this work, we extend this analysis to higher orders, with precision control of both the polarized decay and the polarized production of the different helicity states.

The paper is organized as follows.
Sec.~\ref{sec:rest} sets up the rest-frame correlators of polarized decays and their precision computation.
Sec.~\ref{sec:boost} constructs the polarization-weighted boost kernels and the boosted decay spectra.
Sec.~\ref{sec:production} combines the polarized production rates for $pp\to V+X$ and $e^+e^-\to W^+W^-$ with the boosted spectra and gives our precision predictions.
Sec.~\ref{sec:bump} carries out Monte Carlo studies of electroweak bosons in untagged jet samples, and Sec.~\ref{sec:conclusions} concludes.

\section{Polarized decays at rest}
\label{sec:rest}

We first consider the decay $V\to q\bar q'$, $V=W^\pm,Z$, in its rest frame, and measure the EEC on its decay products. 
We neglect quark masses in the decay and boost formulae.
The EEC distribution between the helicity states $\lambda,\lambda' = \pm 1,0$, quantized along the boson direction, is described by the correlation matrix
\begin{align}\label{eq:wdef}
\Wcal_{\lambda\lambda'}(\vartheta,\varphi,\Phi;z)
\equiv\frac{16\pi^2}{m_V^2}
\langle V,\lambda|\Ecal(\vec n_1)\Ecal(\vec n_2)|V,\lambda'\rangle\,,
\end{align}
where the matrix element of an operator $\mathcal O$ between the source states $|V,\lambda\rangle$ is defined as
\begin{align}\label{eq:state}
\langle V,\lambda|\mathcal O|V,\lambda'\rangle\equiv\frac{\epsilon^{\mu*}_{\lambda}\epsilon^{\nu}_{\lambda'}}{\mathcal N_V}\int\dd^4x\,e^{iq\cdot x}\langle0|J^{V\dagger}_\mu(x)\,\mathcal O\,J^V_\nu(0)|0\rangle\,,
\end{align}
with $q^\mu=(m_V,\vec0)$ and
\begin{align}\label{eq:statenorm}
\mathcal N_V\equiv\frac13\Big(-g^{\mu\nu}+\frac{q^\mu q^\nu}{m_V^2}\Big)\int\dd^4x\,e^{iq\cdot x}\langle0|J^{V\dagger}_\mu(x)J^V_\nu(0)|0\rangle\,,
\end{align}
which is proportional to the total hadronic decay rate, since the tensor in Eq.~\eqref{eq:statenorm} is the polarization sum $\sum_\lambda\epsilon^{\mu*}_\lambda\epsilon^\nu_\lambda$. This gives $\langle V,\lambda|V,\lambda'\rangle=\delta_{\lambda\lambda'}$, with the same normalization for all helicities.
The current $J^V_\mu$ couples $V$ to quarks. For example, $J^{W^+}_\mu=\bar u\gamma_\mu P_Ld$ up to the Cabibbo–Kobayashi–Maskawa (CKM) factor, with the conjugate current for $W^-$.

\begin{figure}[t]
\centering
\includegraphics[width=0.9\columnwidth]{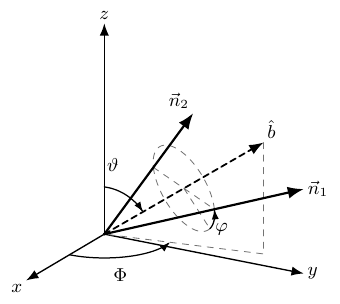}
\caption{Rest-frame detector geometry for
$R=R_z(\Phi)R_y(\vartheta)R_z(\varphi)$.
The $z$ axis is the helicity axis and the boost axis used in
Sec.~\ref{sec:boost}. The $x$ axis lies in the production plane.
The bisector $\hat b=(\vec n_1+\vec n_2)/|\vec n_1+\vec n_2|$
has polar angle $\vartheta$ and azimuth $\Phi$.
The internal angle $\varphi$ rotates the pair about $\hat b$,
measured from $R_z(\Phi)R_y(\vartheta)\hat x$ in the transverse
circle. All detector directions and angles shown are defined before boosting.
}
\label{fig:restangles}
\end{figure}

The dependence on the pair of detector directions $\vec{n}_1$ and $\vec{n}_2$ can be parameterized by the opening-angle variable $z=(1-\vec n_1\cdot\vec n_2)/2$ and the three Euler angles $R=(\vartheta,\varphi,\Phi)$ specifying the orientation of the pair, namely the polar angle $\vartheta$ of its bisector $\hat b=(\vec n_1+\vec n_2)/|\vec n_1+\vec n_2|$ with respect to the helicity axis, the rotation $\varphi$ of the pair about $\hat b$, and the azimuth $\Phi$ of $\hat b$ measured from the production plane, as illustrated in Fig.~\ref{fig:restangles}.
The two detector solid angles factorize into the opening angle and a normalized orientation measure,
\begin{align}\label{eq:orientationmeasure}
\dd\Omega_{\vec n_1}\dd\Omega_{\vec n_2}&=16\pi^2\,\dd z\,\frac{\dd\Phi}{2\pi}\,
\frac{\dd\cos\vartheta}{2}\,\frac{\dd\varphi}{2\pi}\,.
\end{align} 
We also use the shorthand notation $\dd R = \frac{\dd\Phi}{2\pi}\,
\frac{\dd\cos\vartheta}{2}\,\frac{\dd\varphi}{2\pi}$ for the orientation measure, which is normalized to unity, $\int\dd R=1$.
Since the decay products carry total energy $m_V$ and the states are orthonormal, the energy sum rule $\int\dd\Omega_{\vec n_1}\dd\Omega_{\vec n_2}\,\langle V,\lambda|\Ecal(\vec n_1)\Ecal(\vec n_2)|V,\lambda'\rangle=m_V^2\,\delta_{\lambda\lambda'}$ together with Eq.~\eqref{eq:orientationmeasure} gives $\int_0^1\dd z\int\dd R\,\Wcal_{\lambda\lambda}(R;z)=1$ for each helicity, which fixes the prefactor in Eq.~\eqref{eq:wdef}.

The laboratory frame is reached from the rest frame by a boost along the boson momentum, and we take this boost direction as the $z$ axis, the helicity axis of Fig.~\ref{fig:restangles}.
The $x$ axis lies in the production plane, which in the rest frame is spanned by the boost axis and the beam direction, so that the azimuth $\Phi$ of the bisector is measured from the $x$ axis, up to a sign convention for $x$ that affects only the phases of the off-diagonal entries below.
With two detectors we can complete the bisector into the right-handed triad $(\hat e_1,\hat e_2,\hat b)$ attached to the pair, with $\hat e_1=(\vec n_1-\vec n_2)/|\vec n_1-\vec n_2|$ and $\hat e_2=\hat b\times\hat e_1$, and $R=R_z(\Phi)R_y(\vartheta)R_z(\varphi)$ is the rotation that carries the rest-frame axes $(\hat x,\hat y,\hat z)$ into $(\hat e_1,\hat e_2,\hat b)$, and thus also the natural label for the orientation of the pair. Rotational covariance and the symmetries of the decay tensor in Eq.~\eqref{eq:wdef} reduce the nine complex entries of $\Wcal(R;z)\equiv \Wcal(\vartheta,\varphi,\Phi;z)$ to three real structure functions of $z$ alone, each multiplying an operator built from the triad,
\begin{align}\label{eq:Wrotatedops}
\Wcal(R;z)&=\Wcal^{(0,0)}(z)\,\mathbf 1-\Wcal^{(2,0)}(z)\big[3(\hat b\cdot\vec J)^2-2\big]\nonumber\\
&\quad-\tfrac32\,\Wcal^{(2,2)}(z)\big[(\hat e_1\cdot\vec J)^2-(\hat e_2\cdot\vec J)^2\big]\,,
\end{align}
where $\vec J$ denotes the spin-1 generators, and all dependence on the orientation $R$ sits in the operators through the triad. The superscript labels $(j,m)$ come from the angular momentum addition of the bra and ket spins, $\mathbf 1\otimes\mathbf 1=\mathbf 0\oplus\mathbf 1\oplus\mathbf 2$, with $j$ the rank of the operator and $m$ its spin projection along the bisector $\hat b$.

Our decomposition of the decay tensor allows us to evaluate each of its helicity entries in terms of $\Wcal^{(0,0)}$, $\Wcal^{(2,0)}$, and $\Wcal^{(2,2)}$. Using the spin-1 matrices $J_i$ in the helicity basis defined with $\epsilon_\pm=\mp(\hat x\pm i\hat y)/\sqrt2$ and $\epsilon_0=\hat z$, the diagonal entries read
\begin{align}\label{eq:wdiag}
\Wcal_{00}&=\Wcal^{(0,0)}
 +\Wcal^{(2,0)}(3\cos^2\vartheta-1)\nonumber\\
&\quad+\tfrac32\Wcal^{(2,2)}\sin^2\vartheta\cos2\varphi\,,\nonumber\\
\Wcal_{\pm\pm}&=\tfrac12\big(3\Wcal^{(0,0)}-\Wcal_{00}\big)\,,
\end{align}
and the off-diagonal entries read
\begin{align}\label{eq:woffdiagrest}
\Wcal_{+-}&=-\tfrac32\,e^{-2i\Phi}\Big[\Wcal^{(2,0)}\sin^2\vartheta\nonumber\\
&\hspace{0.8em}+\tfrac12\Wcal^{(2,2)}\big((1+\cos^2\vartheta)\cos2\varphi-2i\cos\vartheta\sin2\varphi\big)\Big]\,,\nonumber\\
\Wcal_{+0}&=-\tfrac{3}{\sqrt2}\,e^{-i\Phi}\sin\vartheta\,\Big[\Wcal^{(2,0)}\cos\vartheta\nonumber\\
&\quad-\tfrac12\Wcal^{(2,2)}\big(\cos\vartheta\cos2\varphi-i\sin2\varphi\big)\Big]\,.
\end{align}
The entry $\Wcal_{-0}=-\Wcal_{+0}^*$ follows from Eq.~\eqref{eq:Wrotatedops}, and the remaining entries follow from Hermiticity, $\Wcal_{\lambda'\lambda}=\Wcal_{\lambda\lambda'}^*$.
We will also find it useful to express the correlation matrix in the bisector frame with $\hat b$ along the helicity $z$ axis and $\hat e_1$ along the $x$ axis, which is the orientation $R=\mathbf 1$ in Eqs.~\eqref{eq:wdiag} and~\eqref{eq:woffdiagrest}. It is given as
\begin{align}\label{eq:wbisector}
\Wcal^{\rm bi}_{00}&=\Wcal^{(0,0)}+2\,\Wcal^{(2,0)}\,,&\Wcal^{\rm bi}_{\pm0}&=0\,,\nonumber\\
\Wcal^{\rm bi}_{\pm\pm}&=\Wcal^{(0,0)}-\Wcal^{(2,0)}\,,&
\Wcal^{\rm bi}_{+-}&=-\tfrac32\,\Wcal^{(2,2)}\,.
\end{align}

Denoting by $\sigma_{\lambda\lambda'}$ the production cross section formed from the amplitude for helicity $\lambda$ and the conjugate amplitude for helicity $\lambda'$, with $\sigma_{\rm tot}=\sum_\lambda\sigma_{\lambda\lambda}$, we can write the helicity density matrix as $\rho_{\lambda\lambda'} = \sigma_{\lambda \lambda'}/\sigma_{\rm tot}$, which is normalized to $\sum_\lambda\rho_{\lambda\lambda}=1$. In the narrow-width approximation, in which production and decay factorize, the physical energy correlator of the vector boson is then
\begin{align}\label{eq:rhocontraction}
\frac{1}{\sigma_{\rm tot}}\frac{\dd\sigma_\rho}{\dd z\,\dd R} \approx \sum_{\lambda,\lambda'} \rho_{\lambda'\lambda}\,\Wcal_{\lambda\lambda'}(R;z) = \operatorname{Tr}\big[\rho\,\Wcal(R;z)\big]\,.
\end{align}
In particular, for an unpolarized source, $\rho_{\rm unp}= \frac{\mathbf{1}}{3}$, we find
\begin{align}
\frac{1}{\sigma_{\rm tot}}\frac{\dd\sigma_{\rho_{\rm unp}}}{\dd z\,\dd R} = \tfrac13\big(\Wcal_{00}+\Wcal_{++}+\Wcal_{--}\big) =\Wcal^{(0,0)}(z)\,.\label{eq:unp}
\end{align}
Therefore, the EEC of an unpolarized source has no dependence on any of the three orientation angles.

In the boosted frame we will be interested only in the opening-angle distribution in $z'$. Since the boost along the helicity axis commutes with rotations about it, we can integrate over $\Phi$ already in the rest frame, where the phases in Eq.~\eqref{eq:woffdiagrest} give
\begin{align}\label{eq:azimuthprojection}
\int_0^{2\pi}\frac{\dd\Phi}{2\pi}\,
\Wcal_{\lambda\lambda'}(\vartheta,\varphi,\Phi;z)
=\delta_{\lambda\lambda'}\Wcal_{\lambda\lambda}(\vartheta,\varphi;z)\,.
\end{align}
This projects out the off-diagonal contributions without assuming that $\rho$ is diagonal. Thus, averaged over $\Phi$, Eq.~\eqref{eq:rhocontraction} reduces to $\sum_\lambda\rho_{\lambda\lambda}\Wcal_{\lambda\lambda}(\vartheta,\varphi;z)$.  Therefore, for the normalized $z'$ distribution, we need precision computations of the diagonal entries $\rho_{\lambda\lambda}$ and $\Wcal_{\lambda\lambda}$ only.  Since $\Wcal_{++}=\Wcal_{--}$ for every orientation, the contraction becomes $\rho_{00}\Wcal_{00}+(1-\rho_{00})\Wcal_{++}$, so the production side enters only through the longitudinal fraction $\rho_{00}$, whose dependence on the boost and computation are discussed in Sec.~\ref{sec:production}, and the decay side through the two diagonal entries given in Eq.~\eqref{eq:wdiag}. The simplest case in which the off-diagonal entries survive, a single detector, is worked out in App.~\ref{app:onepoint}.

Integrating in addition over $\vartheta$ and $\varphi$ averages each direction of the triad over the sphere, with $\int\dd R\,(\hat n\cdot\vec J)^2=\tfrac23\mathbf 1$ for $\hat n=\hat b,\hat e_1,\hat e_2$, so the operators multiplying $\Wcal^{(2,0)}$ and $\Wcal^{(2,2)}$ in Eq.~\eqref{eq:Wrotatedops} drop out and
\begin{align}\label{eq:orientationaverage}
\int\dd R\;\Wcal_{\lambda\lambda'}(R;z)=\delta_{\lambda\lambda'}\,\Wcal^{(0,0)}(z)\,.
\end{align}
An opening-angle measurement in the rest frame therefore depends on $\Wcal^{(0,0)}$ alone for any $\rho$, and the polarization becomes visible in the opening angle only after the boost.

\subsection{Precision computation of \texorpdfstring{$\Wcal$}{W}}

Computing any entry of $\Wcal$ reduces to computing the three structure functions $\Wcal^{(0,0)}$, $\Wcal^{(2,0)}$, and $\Wcal^{(2,2)}$.
As described in detail in Ref.~\cite{Gao:2026xuq} for Higgs decays, the rest-frame distribution in $z$ has three distinct angular regions.
In the collinear region $z\ll1$ the two detectors sit inside one jet. In the wide-angle region $z\sim\tfrac12$ they resolve hard wide-angle radiation, which is first populated at $\mathcal O(\alpha_s)$ by $V\to q\bar qg$. In the back-to-back region $1-z\ll1$ they sit in opposite jets.
The angular scale $\mu_{\rm EEC}\sim m_V\sqrt{z}$ in the collinear region and $\mu_{\rm EEC}\sim m_V\sqrt{1-z}$ in the back-to-back region separates these regions from the hard scale $m_V$, and each region calls for a different treatment.

The structure function $\Wcal^{(0,0)}$ is the EEC of an unpolarized source as shown in Eq.~\eqref{eq:unp}.
It carries no orientation dependence, exactly like the EEC of a Higgs decay.
In the wide-angle region the distribution is governed by the hard scale and is obtained from fixed-order perturbation theory, known analytically at NLO~\cite{Dixon:2018qgp} and numerically at NNLO~\cite{Catani:1996jh, Catani:1996vz, Gehrmann-DeRidder:2014hxk, Aveleira:2025svg, DelDuca:2016ily, Tulipant:2017ybb} for vector boson decays. 
In the collinear region the cumulant of the correlator factorizes into a hard function, which describes the production of a fragmenting parton with energy fraction $x$, and a jet function, which describes the correlations generated by collinear splittings inside the jet~\cite{Dixon:2019uzg,Lee:2022uwt,Lee:2024icn},
\begin{align}\label{eq:collfact}
\int_0^{z}\dd \zeta\,\Wcal^{(0,0)}(\zeta)&\overset{z\ll1}{=}\int_0^1\dd x\,x^2\;\vec H^{(0,0)}(x,\mu)\nonumber\\
&\qquad\cdot\vec{\Jcal}^{(0)}\Big(\ln\frac{z\,x^2m_V^2}{\mu^2},\mu\Big)+\mathcal O(z)\,,
\end{align}
where $\vec H^{(0,0)}=(H_q^{(0,0)},H_g^{(0,0)})$ and $\vec{\Jcal}^{(0)}=(\Jcal_q^{(0)},\Jcal_g^{(0)})$ are vectors in flavor space, and the weight $x^2$ reflects the two energy weights of the correlator. The meaning of the superscripts $(0,0)$ and $(0)$ on the hard and jet functions is explained below.
Differentiating with respect to $z$ turns the logarithms of the jet function into the classical $1/z$ scaling of the spectrum and its single-logarithmic corrections, $\alpha_s^n\ln^{n-1}(1/z)/z$, which are resummed by the renormalization group evolution driven by the twist-two spin-three anomalous dimensions.
This resummation is available at NNLL~\cite{Dixon:2019uzg}.
In the back-to-back region the distribution is governed by Sudakov double logarithms of $1-z$ and factorizes~\cite{Moult:2018jzp} into a hard function, two collinear functions describing dijets, and a soft function describing soft recoil, and is currently resummed at N$^4$LL~\cite{COLLINS1981381, deFlorian:2004mp, Moult:2018jzp, Ebert:2020sfi, Moult:2022xzt, Duhr:2022yyp}. The transitions between the regions are obtained by matching the resummed and fixed-order results. We use scale variations to estimate the uncertainties from missing higher-order corrections. In the collinear resummation, we vary the hard and jet scales by a factor of $2$. In the back-to-back resummation, we adopt the 35 variations of Refs.~\cite{Stewart:2013faa,Gao:2024wcg}. In the fixed-order region, we only vary the hard scale. The hard scales in all three regions are varied in a correlated way. Further modifications from $\Lambda_{\rm QCD}$ power corrections~\cite{Korchemsky:1999kt,Schindler:2023cww,Lee:2024esz,Chen:2024nyc,Chen:2026hmd} and from the confinement region are discussed in Refs.~\cite{Lee:2025okn,Chang:2025kgq,Kang:2025zto}.

For the structure functions $\Wcal^{(2,0)}$ and $\Wcal^{(2,2)}$, the same regions appear.
In the wide-angle region $\Wcal^{(2,0)}$ and $\Wcal^{(2,2)}$ are fixed-order quantities like $\Wcal^{(0,0)}$, but computed with the appropriate projection operators to project them out from Eq.~\eqref{eq:Wrotatedops}.
In the wide-angle region, we can express $\Wcal^{(2,m)}$ in terms of $\Wcal^{(0,0)}$ through the ratios~\cite{Riembau:2025isw}
\begin{align}\label{eq:aratios}
a^{(j,m)}(z)\equiv\frac{\Wcal^{(j,m)}(z)}{\Wcal^{(0,0)}(z)}\,,
\qquad (j,m)=(2,0),(2,2)\,.
\end{align}
In the back-to-back region the leading-power factorization allows only two structures compatible with the symmetries of the pair,
\begin{align}\label{eq:b2bdecomp}
\Wcal(R;z)\;\xrightarrow{\,z\to1\,}\;&F(z)\,(\hat e_1\cdot\vec J)^2\nonumber\\
&+G(z)\,\big[(\hat b\cdot\vec J)^2-(\hat e_2\cdot\vec J)^2\big]\,,
\end{align}
where $\hat e_1 \propto \vec{n}_1 - \vec{n}_2$ is now the dijet axis and $\hat b \propto \vec n_1+\vec n_2$ the direction of the transverse-momentum imbalance.
The function $F(z)$ depends on the hard vertex of $V\to q\bar q$ summed over the quark helicities, convoluted with the unpolarized TMD fragmentation functions and the soft function, and it carries the Sudakov logarithms of $\Wcal^{(0,0)}$.
The function $G(z)$ depends on the hard vertex describing the quadrupole about the dijet axis generated by the transverse-spin correlation of the quark and the antiquark.
It arises from the interference of the $q_L\bar q_R$ and $q_R\bar q_L$ amplitudes convoluted with the Collins fragmentation functions of the two jets, and it sets the size of the azimuthal modulation of the imbalance direction $\hat b$ about $\hat e_1$. The interference is proportional to the product $g_Lg_R$ of the couplings of $V$ to left- and right-handed quarks, so $G$ vanishes identically for the $W$, which couples only to left-handed quarks.
Comparison between Eq.~\eqref{eq:b2bdecomp} and Eq.~\eqref{eq:Wrotatedops} gives
\begin{align}\label{eq:b2bFG}
\Wcal^{(0,0)}|_{z\to 1}&=\tfrac23F\,,\qquad
\Wcal^{(2,0)}|_{z\to 1}=\tfrac16F-\tfrac12G\,,\nonumber\\
\Wcal^{(2,2)}|_{z\to 1}&=-\tfrac13F-\tfrac13G\,.
\end{align}
For the energy-weighted correlator, however, each jet contributes to $G$ only through the first transverse-momentum moment of its Collins function summed over all hadrons, which vanishes by the Sch\"afer--Teryaev sum rule~\cite{Schafer:1999kn} as noted in Ref.~\cite{Kang:2023big}.
Hence $G(z)=0$ in the perturbative back-to-back region, $1-z\gg\Lambda_{\rm QCD}^2/m_V^2$. Only the intrinsic transverse-momentum dependence of the Collins functions can generate a nonzero $G$, confined to the nonperturbative endpoint $1-z\lesssim\Lambda_{\rm QCD}^2/m_V^2$, a small effect that we neglect. The back-to-back decay matrix is then $F(z)(\hat e_1\cdot\vec J)^2$ to all orders at leading power.
Eq.~\eqref{eq:b2bFG} then fixes the leading-power back-to-back values of the ratios in Eq.~\eqref{eq:aratios},
\begin{align}\label{eq:b2bratios}
a^{(2,0)}\big|_{z\to1}=\tfrac14\,,\qquad a^{(2,2)}\big|_{z\to1}=-\tfrac12\,.
\end{align}
Thus, in both the wide-angle and back-to-back regions, where the ratios carry no large logarithms, we are able to express $\Wcal^{(2,m)}$ in terms of the ratios $a^{(2,m)}$ and $\Wcal^{(0,0)}$, and at leading power the back-to-back resummation of $\Wcal^{(0,0)}$ carries over to $\Wcal^{(2,m)}$.

In the collinear region, the factorization in Eq.~\eqref{eq:collfact} generalizes to
\begin{align}\label{eq:collfactpol}
\int_0^{z}\dd \zeta\,\Wcal^{(j,m)}(\zeta)&\overset{z\ll1}{=}\int_0^1\dd x\,x^2\;\vec H^{(j,m)}(x,\mu)\nonumber\\
&\qquad\cdot\vec{\Jcal}^{(m)}\Big(\ln\frac{z\,x^2m_V^2}{\mu^2},\mu\Big)+\mathcal O(z)\,.
\end{align}
Here $\vec H^{(j,m)}$ is the $(j,m)$ multipole of the partonic coefficient functions for producing the fragmenting parton from the polarized source, decomposed as in Eq.~\eqref{eq:Wrotatedops} with the jet direction in place of the bisector.
$\vec{\Jcal}^{(m)}$ is the jet function for spin projection $m$ along the jet axis $\hat b$, also called the
transverse spin since it is the charge under rotations about the jet axis, and the same $m$ appears in both factors because these rotations are a symmetry of the collinear limit.

For $m=0$ the jet functions in Eq.~\eqref{eq:collfactpol} are the ordinary EEC jet functions of Eq.~\eqref{eq:collfact} for both $j=0$ and $j=2$, and all dependence on the vector boson spin sits in the hard functions $\vec H^{(0,0)}$ and $\vec H^{(2,0)}$. These jet functions are known to NNLO~\cite{Dixon:2019uzg}.
Like the decay matrix, the hard function is a matrix in the vector boson spin space. Since the hard function produces the fragmenting parton, its spin components are naturally expressed in the frame with the parton direction as the helicity $z$ axis. In the collinear limit, this also corresponds to the bisector frame in Eq.~\eqref{eq:wbisector}. 
The diagonal entries in this frame are the rates for producing the fragmenting parton with energy fraction $x$ from a vector boson that is longitudinal or transverse along the parton direction.
These are the longitudinal and transverse partonic coefficient functions of Refs.~\cite{Nason:1993xx,Rijken:1996ns,Rijken:1996vr,Mitov:2006wy,He:2025hin}, normalized such that the longitudinal entry is $\vec C_L$ and each transverse entry is $\tfrac12\vec C_T$, since $\vec C_T$ counts both transverse helicities.
Eq.~\eqref{eq:collfactpol} therefore factorizes $\Wcal^{\rm bi}_{00}$ with $\vec C_L$ and $\Wcal^{\rm bi}_{\pm\pm}$ with $\tfrac12\vec C_T$.
Inverting Eq.~\eqref{eq:wbisector} for the hard functions gives
\begin{equation}\label{eq:H20}
\vec H^{(0,0)}=\frac13\big(\vec C_T+\vec C_L\big),\quad
\vec H^{(2,0)}=-\frac12\,\vec H^{(0,0)}+\frac12\,\vec C_L\,,
\end{equation}
where both are convolved with the $m=0$ jet function $\vec{\Jcal}^{(0)}$.

For $m=2$, the hard function $\vec{H}^{(2,2)}=(0,H_g^{(2,2)})$ in Eq.~\eqref{eq:collfactpol} is related to the entry $\Wcal_{+-}^{\rm bi}$ of Eq.~\eqref{eq:wbisector}, and thus describes the interference of the vector boson helicities $+1$ and $-1$ along the collinear direction $\hat{b}$. Angular momentum conservation about this axis transfers the difference $\lambda-\lambda' = 2$ to the gluon, so the hard function is the interference of the amplitude for producing the gluon with helicity $+1$ from a vector boson with helicity $+1$ and the amplitude for producing it with helicity $-1$ from a vector boson with helicity $-1$. Since quark helicities can differ by at most one unit,
the $m=2$ channel is purely gluonic at leading
twist~\cite{Chen:2021gdk}. In vector-boson decay, the
corresponding hard function first arises at
$\mathcal O(\alpha_s)$ from $V\to q\bar qg$, with the
gluon as the fragmenting parton.
The label $m=2$ is also called the transverse spin, as it is the eigenvalue,  under rotations about the direction $\hat{b}$, of the twist-two gluon light-ray operator whose matrix element in the vector boson state gives the hard function~\cite{Chen:2021gdk}, and these rotations act only on the plane transverse to $\hat b$. This helicity interference is the linear polarization of the gluon, which the jet function $\Jcal^{(2)}_g$ converts into the $\cos2\varphi$ modulation of the pair about $\hat b$ in Eq.~\eqref{eq:wdiag}.

\begin{figure}[t]
\centering
\includegraphics[scale=0.5]{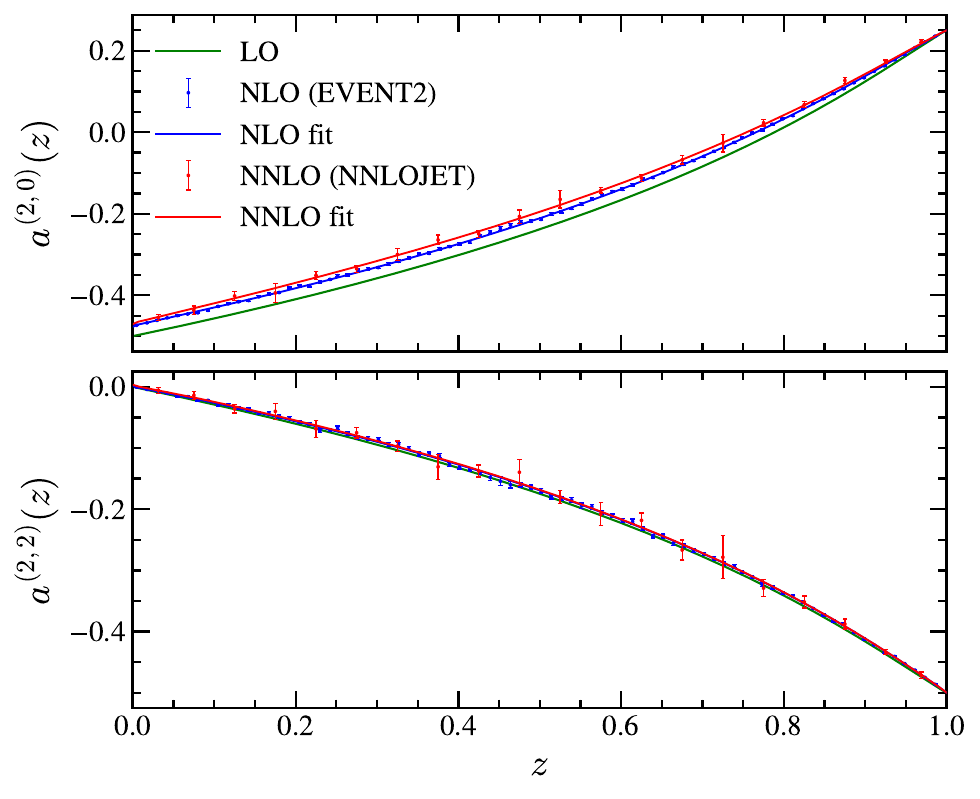}
\caption{\label{fig:ratios}
Rest-frame ratios $a^{(2,0)}$ and $a^{(2,2)}$ of Eq.~\eqref{eq:aratios} at LO, NLO, and NNLO, entering Eq.~\eqref{eq:restinputprescription}, for $\alpha_s=0.118$.
The green curves are the analytic LO results, first computed in Ref.~\cite{Riembau:2025isw}; the blue points are the NLO results computed using \textsc{Event2}~\cite{Catani:1996jh,Catani:1996vz},
and the blue curves are their smooth fits. The red points show the NNLO results computed using \textsc{Nnlojet}~\cite{NNLOJET:2025rno}, and the red curves are their smooth fits.
The NNLO calculation uses roughly $6.4\times10^5$ RR,
$1.1\times10^5$ RV, and $10^3$ VV runs, corresponding to about $6\times10^9$ sector-wise phase-space evaluations. The vertical error bars show Monte Carlo statistical uncertainties. We use the NNLO fitted ratios in our predictions.
}
\end{figure}

This modified factorization structure in the collinear region implies that, unlike in the wide-angle and back-to-back regions, a fixed-order ratio multiplying the resummed $\Wcal^{(0,0)}$ does not reproduce the collinear resummation of $\Wcal^{(2,m)}$, because the hard function for $m=0$, and both the hard and jet functions for $m=2$, are not proportional to those of $\Wcal^{(0,0)}$.
Therefore, the collinear region has to be resummed separately and matched. Putting everything together, our precision computation of $\Wcal^{(2,m)}(z)$ is given by
\begin{align}\label{eq:restinputprescription}
\Wcal^{(2,m)}(z)&=a^{(2,m)}(z)\,\Wcal^{(0,0)}_{\rm FO+b2b}(z)\nonumber\\
&\quad+\Big(\Wcal^{(2,m)}_{\rm coll}-\big[a^{(2,m)}\,\Wcal^{(0,0)}_{\rm FO}\big]_{\rm coll}\Big)(z)\,,
\end{align}
where $a^{(2,m)}$ is the ratio of Eq.~\eqref{eq:aratios} at fixed order, $\Wcal^{(0,0)}_{\rm FO+b2b}$ is the N$^4$LL back-to-back resummation matched to the NNLO fixed-order prediction obtained from \textsc{CoLoRFulNNLO}~\cite{DelDuca:2016ily,Tulipant:2017ybb}, $\Wcal^{(2,m)}_{\rm coll}$ is the collinear resummation of Eq.~\eqref{eq:collfactpol}, and $[\cdots]_{\rm coll}$ is the leading-power collinear limit, which removes the overlap of the two terms.
In the collinear region the first term and the subtracted term cancel at leading power, leaving $\Wcal^{(2,m)}_{\rm coll}$, while in the wide-angle and back-to-back regions the second term lies beyond the fixed-order accuracy of the first.
The same expression applies to $m=0$ and $m=2$, and each ingredient can be improved independently.
The ratios $a^{(2,m)}$ start at $\mathcal O(\alpha_s^0)$, and these leading ratios were first computed in Ref.~\cite{Riembau:2025isw}. We compute their $\mathcal O(\alpha_s)$ and $\mathcal{O}(\alpha_s^2)$ corrections from the structure functions through $\mathcal O(\alpha_s^2)$ with \textsc{Event2}~\cite{Catani:1996jh,Catani:1996vz} and through $\mathcal O(\alpha_s^3)$ with \textsc{Nnlojet}~\cite{NNLOJET:2025rno}, and present the resulting coefficients $a^{(2,m)}$ in Fig.~\ref{fig:ratios} at different orders. 
With the ratios and structure function $\Wcal^{(0,0)}$ both at NNLO, Eq.~\eqref{eq:restinputprescription} achieves the NNLO accuracy away from the collinear region.

To carry out the computation of $\Wcal^{(2,m)}(z)$ also to NNLL accuracy in the collinear region, we need the hard functions $\vec H^{(2,m)}(x,\mu)$ and, for $m=2$, the jet function $\vec{\mathcal{J}}^{(2)}$. For $\vec H^{(2,0)}(x,\mu)$, the only ingredient beyond those of $\vec H^{(0,0)}(x,\mu)$ is the longitudinal coefficient function $\vec C_L$ of Eq.~\eqref{eq:H20}, known already through $\mathcal O(\alpha_s^3)$~\cite{Nason:1993xx,Rijken:1996vr,Rijken:1996ns,Mitov:2006wy,He:2025hin}, so the collinear-resummed $\Wcal^{(2,0)}_{\rm coll}$ is available at the same NNLL accuracy as the collinear resummation of $\Wcal^{(0,0)}$. For $m=2$, both $\vec H^{(2,2)}(x,\mu)$ and $\vec{\mathcal{J}}^{(2)}$ are only known at their first order, which is not sufficient for the NNLL accuracy of the collinear resummation of $\Wcal^{(2,2)}$. The jet function $\Jcal^{(2)}_g$ is known at $\mathcal O(\alpha_s)$, together with the one-loop anomalous dimension of the transverse-spin-two gluon operator~\cite{Chen:2021gdk}. The $\mathcal{O}(\alpha_s)$ gluon helicity-interference structure entering $H_g^{(2,2)}$ can be extracted from the massless-quark limit of the gluon–beam-plane polarization calculation in Ref.~\cite{Groote:1998qt}.
With these ingredients $\Wcal^{(2,2)}$ could be resummed only through its leading tower $\alpha_s^n\ln^{n-2}(1/z)/z$, which is NLL in the counting of $\Wcal^{(0,0)}$ and thus short of NNLL precision. However, the collinear region of $\Wcal^{(2,2)}$ does not affect the precision of measurements that are sensitive only to the opening angle $z$ or $z'$. In the rest frame, averaging over the orientation $R$ of the pair removes the dependence on $\Wcal^{(2,0)}$ and $\Wcal^{(2,2)}$ in Eq.~\eqref{eq:Wrotatedops}, so the opening-angle $z$ distribution depends on $\Wcal^{(0,0)}$ alone, as in Eq.~\eqref{eq:orientationaverage}. As we will see in detail in Sec.~\ref{sec:kernels}, in the boosted frame the boost integrates the $\cos2\varphi$ weight of $\Wcal^{(2,2)}$ to zero at leading power in the boosted collinear region $\gamma^2z'\ll1$, so the collinear logarithms of the boosted spectrum receive no contribution from $m=2$, while the collinear tower of $\Wcal^{(2,0)}$ does contribute. Therefore, our boosted distribution of the opening angle $z'$ retains NNLL accuracy in the collinear region. 

\section{Boosted polarized decays}
\label{sec:boost}

From now on, we consider the boosted EEC differential in the opening angle $z'$ without measuring the global azimuth $\Phi$.
As shown in Eq.~\eqref{eq:azimuthprojection}, the $\Phi$ integral removes all off-diagonal contributions. Measuring in addition the orientation of the detectors about the boost axis gives access to the off-diagonal entries of $\rho$, as we illustrate for a single detector in App.~\ref{app:onepoint}.
We therefore need only the diagonal rest-frame components $\Wcal_{\lambda\lambda}$ in Eq.~\eqref{eq:wdiag}, without assuming that the production density matrix is diagonal.
We take the boost along the $z$ axis shown in Fig.~\ref{fig:restangles}.

For a boson produced with boost $\gamma$, the boosted analogue of Eq.~\eqref{eq:rhocontraction} for the opening-angle distribution therefore involves only the diagonal entries and reads, in the narrow-width approximation,
\begin{align}\label{eq:rhocontractionboost}
\frac{1}{\sigma_{\rm tot}}\frac{\dd\sigma_\rho}{\dd z'}&\approx\sum_{\lambda}\rho_{\lambda\lambda}\,\Wcal^{\rm boost}_{\lambda\lambda}(z';\gamma)\nonumber\\
&=\rho_{00}\,\Wcal^{\rm boost}_{00}(z';\gamma)+(1-\rho_{00})\,\Wcal^{\rm boost}_{++}(z';\gamma)\,,
\end{align}
where $\Wcal^{\rm boost}_{\lambda\lambda}(z';\gamma)$ is the opening-angle spectrum of a boosted boson with helicity $\lambda$, constructed below, and the second line uses $\Wcal^{\rm boost}_{++}=\Wcal^{\rm boost}_{--}$, which follows from $\Wcal_{++}=\Wcal_{--}$.
How the diagonal entries $\rho_{\lambda\lambda}$, now functions of the boost, are computed is discussed in Sec.~\ref{sec:production}.

The boost supplies the external direction needed to turn the orientation dependence of the rest-frame correlator into opening-angle dependence.
Under a   Lorentz boost $\Lambda$, the null detector directions satisfy $ n_i'=\frac{\Lambda n_i}{\lambda_i}$, with $n_i^\mu=(1,\vec n_i)$, $n_i^{\prime\mu}=(1,\vec n_i')$ and positive Doppler factors $\lambda_i$. 
In terms of the orientation angles, the Doppler factors for a boson moving along $+\hat z$ are given as
\begin{equation}\label{eq:doppler}
\lambda_{1,2} = \gamma + \sqrt{\gamma^2-1}\big(\cos\tfrac{\theta}{2}\cos\vartheta\mp\sin\tfrac{\theta}{2}\sin\vartheta\cos\varphi\big),
\end{equation}
with $z = (1-\cos\theta)/2$.
The energy-flow operator and detector solid angle transform as~\mbox{\cite{Belitsky:2013ofa,Riembau:2025isw}
}\hskip0pt\begin{align}\label{eq:boosttransformation}
U^\dagger(\Lambda)\Ecal(\vec n_i')U(\Lambda)
 &=\lambda_i^3\Ecal(\vec n_i)\,,\nonumber\\
\dd\Omega_{\vec n_i'}&=\lambda_i^{-2}\dd\Omega_{\vec n_i}\,.
\end{align}
The rest-frame opening angle thus maps to
  \begin{align}\label{eq:zpmap}
z'=\frac{z}{\lambda_1\lambda_2}\,,
\end{align} 
which depends on the   pair orientation  relative to the boost axis   but not on a common rotation about that axis.

With $|V,\lambda;\gamma\rangle=U(\Lambda)|V,\lambda\rangle$, the boosted spectrum of Eq.~\eqref{eq:rhocontractionboost} is defined as
\begin{align}\label{eq:Wdef}
\Wcal^{\rm boost}_{\lambda\lambda}(z';\gamma)
 &=\frac{1}{(\gamma m_V)^2}
 \int\dd\Omega_{\vec n_1'}\dd\Omega_{\vec n_2'}\,
 \delta\Big(z'-\tfrac{1-\vec n_1'\cdot\vec n_2'}{2}\Big)\nonumber\\
 &\quad\times\langle V,\lambda;\gamma|
 \Ecal(\vec n_1')\Ecal(\vec n_2')|V,\lambda;\gamma\rangle\,,
\end{align} 
where the boosted vector boson has energy $E_V = \gamma m_V$, which sets the normalization of the two energy weights.
Using Eqs.~\eqref{eq:wdef} and~\eqref{eq:boosttransformation}, and performing the common-azimuth integral, gives the exact boost map
\begin{align}\label{eq:masterboost}
\Wcal^{\rm boost}_{\lambda\lambda}(z';\gamma)
 &=\frac{1}{\gamma^2}\int_0^1\dd z
 \int\frac{\dd\cos\vartheta}{2}\frac{\dd\varphi}{2\pi}\,
 \lambda_1\lambda_2\nonumber\\
 &\quad\times\delta\Big(z'-\frac{z}{\lambda_1\lambda_2}\Big)
 \Wcal_{\lambda\lambda}(\vartheta,\varphi;z)\,.
\end{align}

By Eq.~\eqref{eq:wdiag}, each diagonal entry $\Wcal_{\lambda\lambda}(\vartheta,\varphi;z)$ is a sum of $\Wcal^{(0,0)}(z)$, $\Wcal^{(2,0)}(z)$, and $\Wcal^{(2,2)}(z)$, multiplied, up to helicity-dependent constants, by the respective elementary angular weights
\begin{equation}\label{eq:weights}
w(\vartheta,\varphi)\in\{1,\ 3\cos^2\vartheta-1,\ \sin^2\vartheta\cos2\varphi\}\,.
\end{equation}
The structure functions $\Wcal^{(j,m)}(z)$ therefore factor out of the orientation integral in Eq.~\eqref{eq:masterboost}, and at fixed $z$ we define the purely kinematic weighted kernels
\begin{equation}\label{eq:Kw}
\Kcal_\gamma^{[w]}(z,z')\equiv\frac{1}{\gamma^2}
\int\frac{\dd^2\Omega}{4\pi}\,\lambda_1\lambda_2\,
\delta\Big(z'-\frac{z}{\lambda_1\lambda_2}\Big)w(\vartheta,\varphi)\,,
\end{equation}
where $\dd^2\Omega/(4\pi)=(\dd\cos\vartheta/2)(\dd\varphi/2\pi)$ and $w(\vartheta,\varphi)$ is only a prescribed angular weight, not a helicity correlator.
The unweighted kernel $\Kcal_\gamma\equiv\Kcal_\gamma^{[1]}$ is the scalar boost kernel of Eq.~\eqref{eq:scalarboost}, which was computed in closed form for boosted Higgs decays in Ref.~\cite{Gao:2026xuq}.
In terms of these kernels, Eq.~\eqref{eq:masterboost} becomes a sum of three one-dimensional convolutions, one for each structure function, which gives the longitudinal spectrum
\begin{align}\label{eq:W00}
\Wcal^{\rm boost}_{00}(z';\gamma)
 &=\int_0^1\dd z\,\Big[
 \Kcal_\gamma^{[1]}(z,z')\Wcal^{(0,0)}(z)\nonumber\\
 &\quad+\Kcal_\gamma^{[3\cos^2\vartheta-1]}(z,z')\Wcal^{(2,0)}(z)\nonumber\\
 &\quad+\tfrac32\Kcal_\gamma^{[\sin^2\vartheta\cos2\varphi]}(z,z')
 \Wcal^{(2,2)}(z)\Big]\,,
\end{align}
and the common transverse spectrum
\begin{align}\label{eq:Wtransverse}
\Wcal^{\rm boost}_{++}(z';\gamma)
 &=\Wcal^{\rm boost}_{--}(z';\gamma)\nonumber\\
 &=\int_0^1\dd z\,\Big[
 \Kcal_\gamma^{[1]}(z,z')\Wcal^{(0,0)}(z)\nonumber\\
 &\quad-\tfrac12\Kcal_\gamma^{[3\cos^2\vartheta-1]}(z,z')\Wcal^{(2,0)}(z)\nonumber\\
 &\quad-\tfrac34\Kcal_\gamma^{[\sin^2\vartheta\cos2\varphi]}(z,z')
 \Wcal^{(2,2)}(z)\Big]\,.
\end{align}
All dependence on the decay dynamics is contained in the three $\Wcal^{(j,m)}$, and the same three boost kernels apply to each helicity.

\subsection{Polarization-weighted boost kernels}
\label{sec:kernels}

The orientation angles $\vartheta,\varphi$ in Eq.~\eqref{eq:Kw} are defined in the rest-frame geometry of Fig.~\ref{fig:restangles}, with $z$ as the boost axis.

Note that in the zero-boost limit ($\gamma\to1$), Eq.~\eqref{eq:Kw} reduces to
\begin{align}
    \Kcal_{\gamma=1}^{[w]}(z,z')=\delta(z'-z)
    \int\!\frac{\dd^2\Omega}{4\pi}\,w(\vartheta,\varphi)\,,
\end{align}
which is $\delta(z'-z)$ for $w=1$, and zero for $w=3\cos^2\vartheta-1$ and $w=\sin^2\vartheta \cos2\varphi$.
This is consistent with Eq.~\eqref{eq:orientationaverage}, which says that in the rest frame, if one only measures the relative angle $z$ between the two detectors, one does not probe polarization.

The unweighted kernel evaluates to the closed form of Ref.~\cite{Gao:2026xuq},
\begin{align}\label{eq:Kscalar}
\Kcal_\gamma(z,z') &= \frac{z^{3/2}\;\Theta\big(\xi(z,z')\big)}{2\pi\gamma^2\sqrt{\gamma^2-1}\; z'^{5/2}\,\sqrt[4]{(1-z)(1-z')}} \nonumber\\
&\hspace{2mm}\times
\begin{cases}
K\big(\sqrt{\xi}\big)\,, & 0\le \xi<1\,,\\[3pt]
\frac{1}{\sqrt{\xi}}\,K\big(1/\sqrt{\xi}\big)\,, & \xi>1\,,
\end{cases}
\end{align}
with $K$ the complete elliptic integral of the first kind and
\begin{align}
\xi(z,z') = \frac{2\sqrt{1-z}\sqrt{1-z'} + z + z' + z z'(\gamma^2-1) - 2}{4\sqrt{1-z}\sqrt{1-z'}}\,.
\end{align}
The kernel is supported on $z'_- \le z' \le z'_+$ for $z \le 1/\gamma^2$, and on $z'_- \le z' \le 1$ for $z > 1/\gamma^2$, where $z'_{\pm} = z\big(\gamma \mp \sqrt{\gamma^2-1}\sqrt{1-z}\big)^{-2}$ and, in the second case, $z'_+$ marks the branch point $\xi=1$ at which the kernel has an integrable logarithmic singularity. 
As $z\to1$ both $z'_\pm$ approach $1/\gamma^2$, squeezing the first branch onto a small region near $z'=1/\gamma^2$, while the second branch becomes the whole interval $1/\gamma^2\approx z_+'\le z'\le1$. The general weighted boost kernels involve complete elliptic integrals $K$, $E$, and $\Pi$.
Full expressions, branch structure, and endpoint limits are collected in App.~\ref{app:kernels}.

We further note that, in the collinear limit, the boost kernels simplify significantly. Writing $u=\cos\vartheta$ and $\kappa=\sqrt{\gamma^2-1}$, Eq.~\eqref{eq:doppler} gives
\begin{align}\label{eq:collinearboostproduct}
\lambda_1\lambda_2
&=(\gamma+\kappa\sqrt{1-z}\,u)^2
-\kappa^2 z(1-u^2)\cos^2\varphi\nonumber\\
&=(\gamma+\kappa u)^2+\mathcal O(z)\,.
\end{align} 
At leading power in $z$, $\lambda_1\lambda_2$ is therefore independent of $\varphi$, so neither the factor $\lambda_1\lambda_2$ nor the opening-angle constraint $\delta(z'-z/\lambda_1\lambda_2)$ in Eq.~\eqref{eq:Kw} depends on $\varphi$, and the $\varphi$ integral acts only on the weight $w(\vartheta,\varphi)$.
The weights $1$ and $3\cos^2\vartheta-1$ of $\Wcal^{(0,0)}$ and $\Wcal^{(2,0)}$ do not depend on $\varphi$, and their kernels map a leading $1/z$ term onto $1/z'$ with the factors $\gamma^{-2}\int_{-1}^{1}\frac{\dd u}{2}\,(\gamma+\kappa u)^2\,w(u)$, which equal $(4\gamma^2-1)/(3\gamma^2)$ and $4(\gamma^2-1)/(15\gamma^2)$.
More generally, a rest-frame collinear term $\ln^k(1/z)/z$ maps onto terms $\ln^j(1/z')/z'$ with $j\leq k$, whose prefactors are orientation averages of powers of $\ln(\lambda_1\lambda_2)$.
The collinear tower of $\Wcal^{(2,0)}$ is therefore not suppressed by the boost, which is why it is resummed at NNLL through the hard function of Eq.~\eqref{eq:H20}.
The weight $\sin^2\vartheta\cos2\varphi$ of $\Wcal^{(2,2)}$ instead integrates to zero over $\varphi$, since $\int_0^{2\pi}\dd\varphi\,\cos2\varphi=0$.
This removes every $\ln^k(1/z')/z'$ term from that channel, and the first surviving terms, which come from the $\varphi$-dependent $\mathcal O(z)$ term in Eq.~\eqref{eq:collinearboostproduct}, are suppressed by a power of $\gamma^2z'$.
The statement holds at fixed boost for $\gamma^2z'\ll1$ and for an opening-angle measurement, which integrates over $\varphi$.
The limited accuracy of $H^{(2,2)}_g$ and $\Jcal^{(2)}_g$ discussed in Sec.~\ref{sec:rest} therefore does not affect the leading-power collinear logarithms of the boosted spectra.

\begin{figure}[t]
\includegraphics[width=\columnwidth]{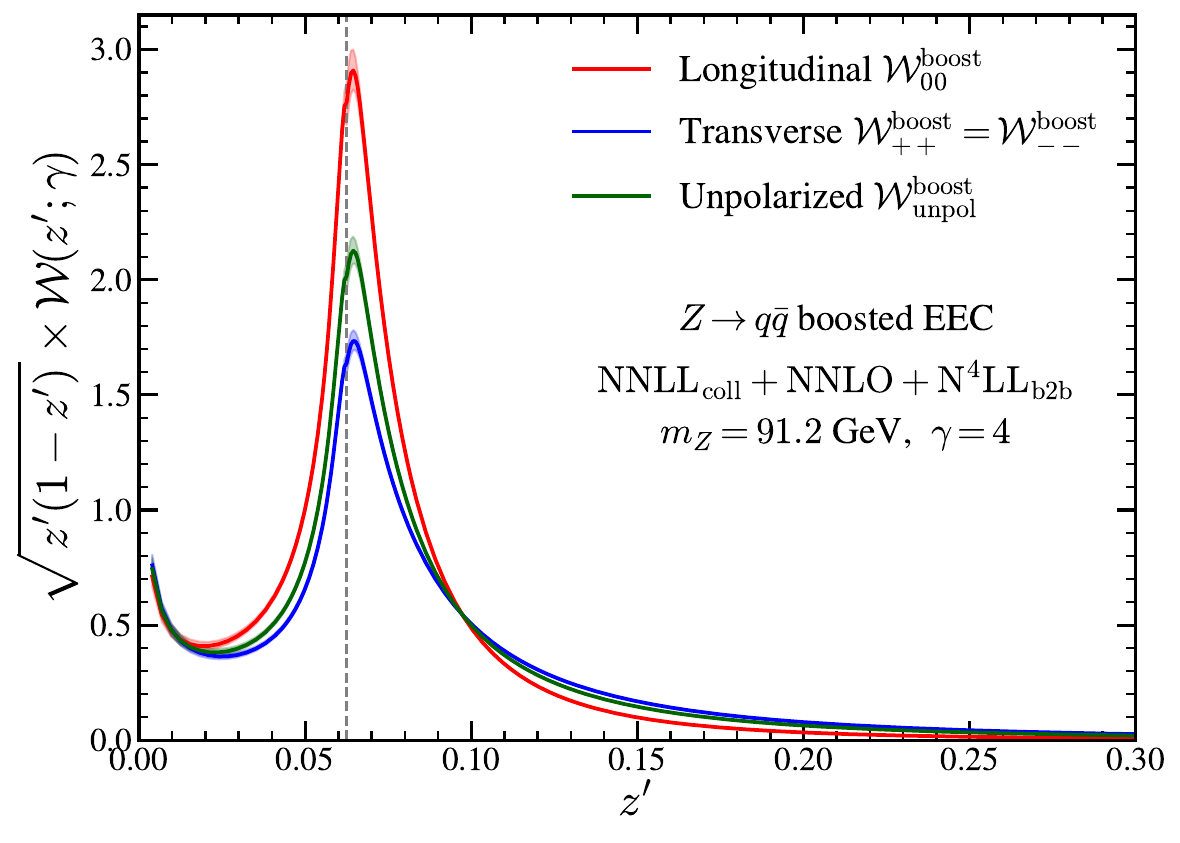}
\caption{\label{fig:LT}
Boosted EEC of a polarized decay $Z\to q\bar q$ at fixed boost $\gamma=4$ and $Q = 91.2$~GeV, at ${\rm NNLL_{coll}} + {\rm N^4LL_{b2b}} + {\rm NNLO}$ accuracy, weighted by $\sqrt{z'(1-z')}$ for display.
Shown are the longitudinal spectrum $\Wcal^{\rm boost}_{00}$, with $\rho_{00}=1$ (red), the transverse spectrum $\Wcal^{\rm boost}_{++}=\Wcal^{\rm boost}_{--}$, with $\rho_{00}=0$ (blue), and the unpolarized spectrum of Eq.~\eqref{eq:trace}, with $\rho_{00}=\rho_{\pm\pm}=1/3$ (green).
The decay peak at $z' \simeq 1/\gamma^2$ is enhanced for longitudinal and depleted for transverse polarization, with a height difference of roughly $70\%$ between the two, making the peak a polarimeter built from angular information alone.
Bands denote perturbative uncertainties.
}
\end{figure}

\subsection{Polarization dependence of the decay peak}
\label{sec:peak}
To study the different polarizations, we set $\rho_{00}=1$ or $\rho_{00}=0$ in Eq.~\eqref{eq:rhocontractionboost}, which selects the longitudinal spectrum $\Wcal^{\rm boost}_{00}$ or the transverse spectrum $\Wcal^{\rm boost}_{++}=\Wcal^{\rm boost}_{--}$, and we compare both with the unpolarized spectrum, obtained with $\rho=\mathbf{1}/3$, 
\begin{align}\label{eq:trace}
\frac{1}{\sigma_{\rm tot}}\frac{\dd\sigma_{\rho_{\rm unp}}}{\dd z'}
&=\tfrac13\big(\Wcal^{\rm boost}_{00}+\Wcal^{\rm boost}_{++}+\Wcal^{\rm boost}_{--}\big)\nonumber\\
&=\int_0^1\dd z\,\Kcal_\gamma(z,z')\Wcal^{(0,0)}(z)\,.
\end{align}
Thus the unpolarized source recovers the scalar boost map of Ref.~\cite{Gao:2026xuq}, applied to its own spin-averaged decay EEC. The boosted spectra also obey the energy and momentum sum rules for each helicity, as shown in App.~\ref{app:kernels}.

The physics of the polarization dependence is transparent at the decay peak.
Rest-frame exact back-to-back configurations, $\propto \delta(1-z)$, map onto $z'=1/[\gamma^2-(\gamma^2-1)\cos^2\theta_*]\geq 1/\gamma^2$, where $\theta_*$ is the rest-frame angle between the pair axis and the boost axis. Pairs perpendicular to the boost axis land at $z'=1/\gamma^2$ and pairs along it remain at $z'=1$, and since $z'$ is stationary at $\theta_*=\pi/2$ the decays pile up at the lower endpoint with an integrable singularity $\propto(\gamma^2z'-1)^{-1/2}$ at $z' = 1/\gamma^2$, Eq.~\eqref{eq:appz1}.
Sudakov broadening of the back-to-back region converts this singularity into the finite decay peak~\cite{Gao:2026xuq}.
A longitudinal $V$ decays preferentially perpendicular to its spin axis ($\propto \sin^2\theta_*$), feeding exactly the orientations that pile up at the peak.
Transverse polarizations ($\propto (1+\cos^2\theta_*)/2$) instead favor alignment with the boost axis, redistributing correlation toward larger $z'$.
As shown in Fig.~\ref{fig:LT}, at $\gamma=4$ the peak height differs by $\sim 70\%$ between purely longitudinal and purely transverse samples. In the deep collinear region $z'\ll1/\gamma^2$, governed by QCD radiation inside each subjet, the two spectra share the same scaling but not the same normalization, since a transverse boson emits its jets preferentially along the boost axis, where the forward jet carries most of the boson energy and the quadratic energy weights of the correlator are largest.

\section{From polarized production to collider spectra}
\label{sec:production}

\subsection{Polarized production rates}
\label{sec:master}

{

The production side of Eq.~\eqref{eq:rhocontractionboost} enters only through the diagonal entries of the density matrix.

Let $\dd\sigma^{\rm prod}_{\lambda\lambda}/\dd\gamma$ denote the production rate of a stable on-shell boson with helicity $\lambda$ along its laboratory momentum, with the other production variables, such as the transverse momentum and rapidity, integrated over the selection at fixed boost factor $\gamma$. We define
\begin{align}\label{eq:productionnormalization}
\frac{\dd\sigma^{\rm prod}}{\dd\gamma}
 &=\sum_{\lambda=0,\pm1}\frac{\dd\sigma^{\rm prod}_{\lambda\lambda}}{\dd\gamma}\,,\nonumber\\
\rho_{\lambda\lambda}(\gamma)
 &=\frac{\dd\sigma^{\rm prod}_{\lambda\lambda}/\dd\gamma}
         {\dd\sigma^{\rm prod}/\dd\gamma}\,,
\qquad \sum_\lambda\rho_{\lambda\lambda}=1\,,
\end{align}
so that the opening-angle EEC distribution requires only the total production rate $\dd\sigma^{\rm prod}/\dd\gamma$ and the longitudinal fraction $\rho_{00}(\gamma)$, defined in the helicity frame for the stated selection.

For a specified hadronic decay channel, write $B_{\rm had}=\Gamma_{\rm had}/\Gamma_V$, with $\Gamma_{\rm had}$ the hadronic partial width of that channel and $\Gamma_V$ the total width. Multiplying Eq.~\eqref{eq:rhocontractionboost} by the production rate gives the energy-weighted cross section
\begin{align}\label{eq:master}
\frac{\dd^2\sigma}{\dd\gamma\,\dd z'}
 &=B_{\rm had}\,\frac{\dd\sigma^{\rm prod}}{\dd\gamma}\Big[\rho_{00}(\gamma)\,\Wcal^{\rm boost}_{00}(z';\gamma)\nonumber\\
 &\quad+\big(1-\rho_{00}(\gamma)\big)\,\Wcal^{\rm boost}_{++}(z';\gamma)\Big]\,.
\end{align}

As an unpolarized reference we use the same production spectrum with $\rho_{00}=1/3$, which gives Eq.~\eqref{eq:trace}. An opening-angle measurement constrains only $\rho_{00}$ and cannot separate $\rho_{++}$ from $\rho_{--}$.

The factor $B_{\rm had}$ cancels in the normalized spectra.
 Production cuts can be included in the polarized rates. Additional cuts on the measured decay products generally require acceptance-dependent boosted spectra.

}

\subsection{\texorpdfstring{Polarized spectra in $pp \to V+X$}{Polarized spectra in pp -> V+X}}
\label{sec:ppprod}

We compute the production density matrix for stable on-shell $V$ production through NNLO in QCD~\cite{Gehrmann-DeRidder:2015wbt,Gehrmann-DeRidder:2017mvr} using \textsc{Nnlojet}~\cite{NNLOJET:2025rno}, differential in the boson transverse momentum $p_T$ and rapidity $y_V$.
The boost distribution for $pp\rightarrow V+X$ then follows as
\begin{equation}
\frac{\dd\sigma}{\dd\gamma}
= \int \dd p_{T}\, \dd y_V\; \frac{\dd\sigma}{\dd p_{T}\,\dd y_V}\;
\delta\Big(\gamma - \frac{\sqrt{p_{T}^2+m_V^2}\,\cosh y_V}{m_V}\Big)\,.
\end{equation}
\textsc{Nnlojet} does not provide stable polarized bosons, so we let them decay leptonically and use the leptons as a polarimeter: $Z\to\ell^+\ell^-$ within $66<m_{\ell\ell}<116$~GeV, and $W^\pm\to\ell^\pm\nu$ within $60<m_{\ell\nu}<100$~GeV. 
We impose no cuts on the individual leptons and extract the polarization fractions from their decay angular distributions in the helicity frame~\cite{Mirkes:1994eb}. Then we use the branching ratios, with the Breit-Wigner propagator inside the mass window, to obtain the production cross-sections with stable polarized bosons. 
Up to NNLO, we use $\sqrt{s}=13.6$~TeV and the NNPDF3.1 NLO parton distributions~\cite{NNPDF:2017mvq} with $\alpha_s(m_Z)=0.118$. For electroweak particles, we use $m_Z=91.1876$~GeV, $\Gamma_Z=2.4952$~GeV, $m_W=80.385$~GeV, $\Gamma_W=2.085$~GeV, and the electroweak couplings in the $G_\mu$ scheme with $G_F=1.1663787\times10^{-5}$~GeV$^{-2}$, in which the weak mixing angle is fixed by the boson masses, $\sin^2\theta_W=1-m_W^2/m_Z^2$, and the electromagnetic coupling by the Fermi constant, $\alpha=\sqrt2G_Fm_W^2\sin^2\theta_W/\pi$.
In the final state, we place a $p_T$ cut for the boson, $p_T>100$~GeV.
The central renormalization and factorization scales are set to the transverse energy of the boson, $\mu_R=\mu_F=E_{T,V}=(m_{\ell\ell}^2+p_{T}^2)^{1/2}$, with $m_{\ell\nu}$ in place of $m_{\ell\ell}$ for the $W$. The scale uncertainty is the envelope of the seven-point variation of $\mu_R$ and $\mu_F$ by factors of two with $1/2\le\mu_R/\mu_F\le2$.

In the implementation, we consider the following $p_T$ bins, 100, 125, 150, 175, 200, 250, 300, 350, 400, 500, 600, and 800~GeV. We approximate $y_V=0$ by the rapidity slice  $|y_V|<0.5$. At $y_V=0$ the boost factor is $\gamma=(1+p_{T}^2/m_V^2)^{1/2}$, and $100<p_{T}<800$~GeV corresponds to $1.48<\gamma<8.83$ for the $Z$ and $1.60<\gamma<10.0$ for the $W$. We then fit the numerical data with quartic polynomials in $\ln\gamma$ and obtain the form for $\rho_{00}(\gamma)$ for each scale choice.

\begin{figure}[t]
\centering
\includegraphics[width=\columnwidth]{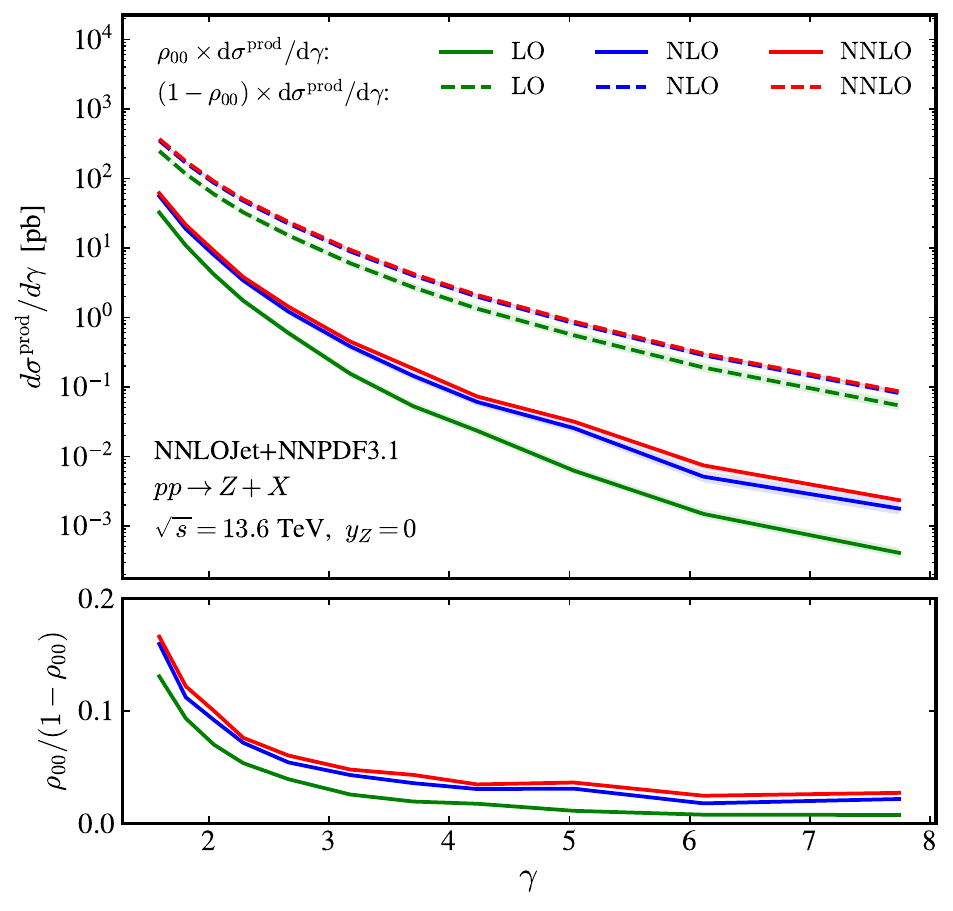}
\caption{
Polarized production spectra  $\dd\sigma/\dd\gamma$ for $pp\to Z+X$ at $\sqrt{s}=13.6$~TeV,   $y_Z = 0$ , $p_T>100$~GeV (stable-$Z$ production), showing the longitudinal rate $\dd\sigma^{\rm prod}_{00}/\dd\gamma$ (solid) and the transverse rate $\sum_\pm\dd\sigma^{\rm prod}_{\pm\pm}/\dd\gamma$ (dashed)  at LO, NLO, and NNLO.
The lower panel shows $\rho_{00}/(1-\rho_{00})=\sigma_{00}/(\sigma_{++}+\sigma_{--})$. The line styles distinguish the longitudinal (solid) and transverse (dashed) production; colors distinguish LO, NLO, and NNLO. The corresponding results for $W^\pm$ production are shown in App.~\ref{app:Wprod}. In \textsc{Nnlojet}, we have to let $Z$ decay and we use a $Z$ lineshape window within $m_{ll}\in[66,116]$ GeV to reconstruct $Z$.
}
\label{fig:LTproduction}
\end{figure}

Fig.~\ref{fig:LTproduction} shows the longitudinal and total transverse boost spectra for $pp\to Z+X$ to NNLO.
High-$p_T$ singly produced $V$ bosons are strongly transverse dominated, with $\rho_{00}/(1-\rho_{00})$ decreasing from ${\approx}14\%$ at $\gamma\approx1.7$ to $2$--$3\%$ at $\gamma\approx7$ for the $Z$, with slightly smaller values for the $W^\pm$ in Fig.~\ref{fig:LTproductionW}.
The NLO correction to the ratio is sizable, up to a factor of a few at large $\gamma$ and driven by the longitudinal channel, but the result stabilizes between NLO and NNLO.
The analogous results for $W^+$ and $W^-$ production are shown in Fig.~\ref{fig:LTproductionW} of App.~\ref{app:Wprod}, with the same line-style convention.

\begin{figure}[t]
\includegraphics[width=\columnwidth]{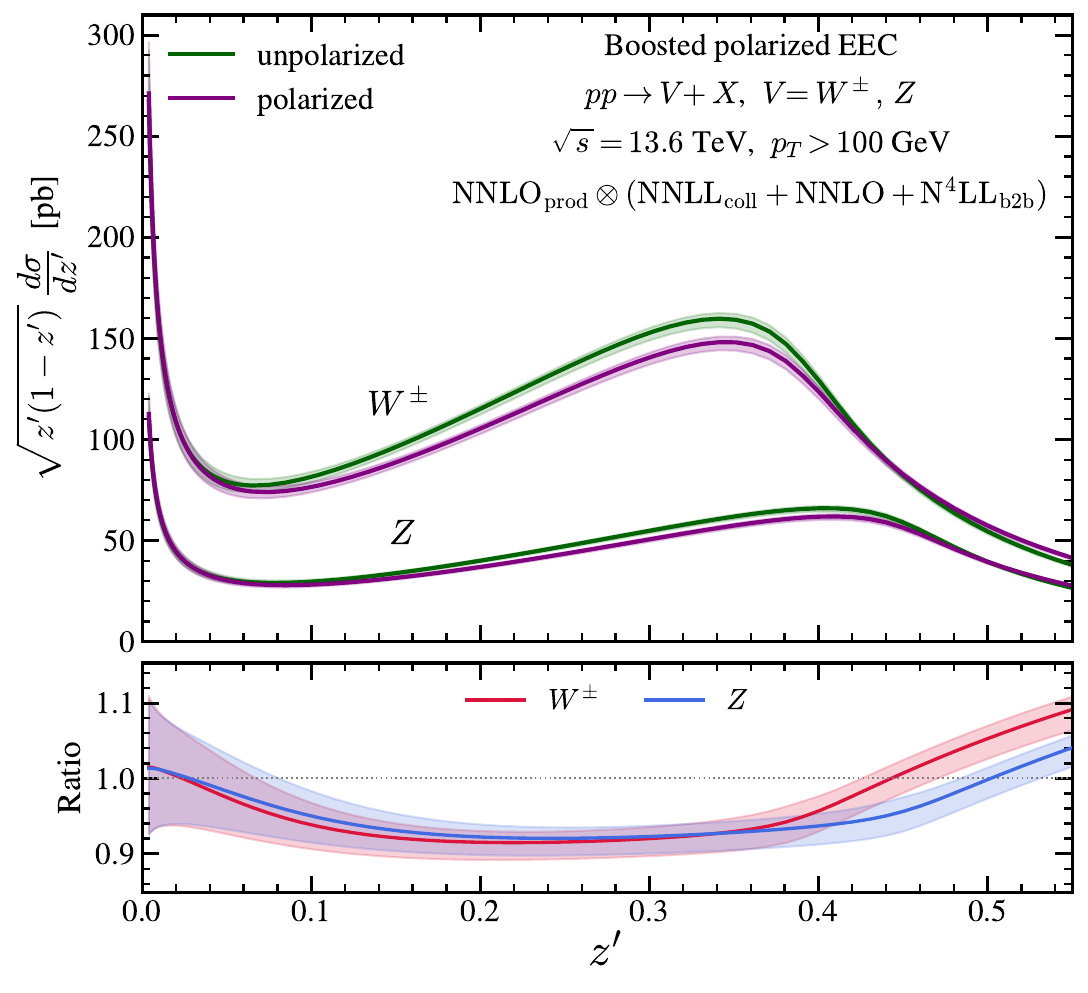}\\[4pt]
\caption{\label{fig:money}
Boosted  EEC  predictions  for   the hadronic decays of $V=W^\pm,Z$ in $pp\to V+X$  at $\sqrt{s}=13.6$~TeV,   $y_V=0$, integrated over   $p_T>100$ ~GeV,   with the  production   and  decay   accuracies indicated in the plot. The physical polarization mixture  (purple)   is  compared with the  unpolarized   reference  (green).
The lower panel shows physical/unpolarized separately for the   $W$ and $Z$ channels. The $W^\pm$ ratio is shifted to smaller $z'$ because the same $p_T$ cut selects more boosted $W$ bosons, $\gamma_{\min}=\sqrt{1+p_T^2/m_V^2}=1.60$ versus $1.48$ for the $Z$.
Accounting for the polarization content of high-$p_T$ $V$ production suppresses the spectrum by up to ${\sim}10\%$, with the effect spread over the $\gamma$-smeared peak region $0.1 \lesssim z' \lesssim 0.4$ and reduced to a percent-level shift at the smallest $z'$.
Bands denote perturbative uncertainties.}
\end{figure}

\begin{figure}[t]
\includegraphics[width=\columnwidth]{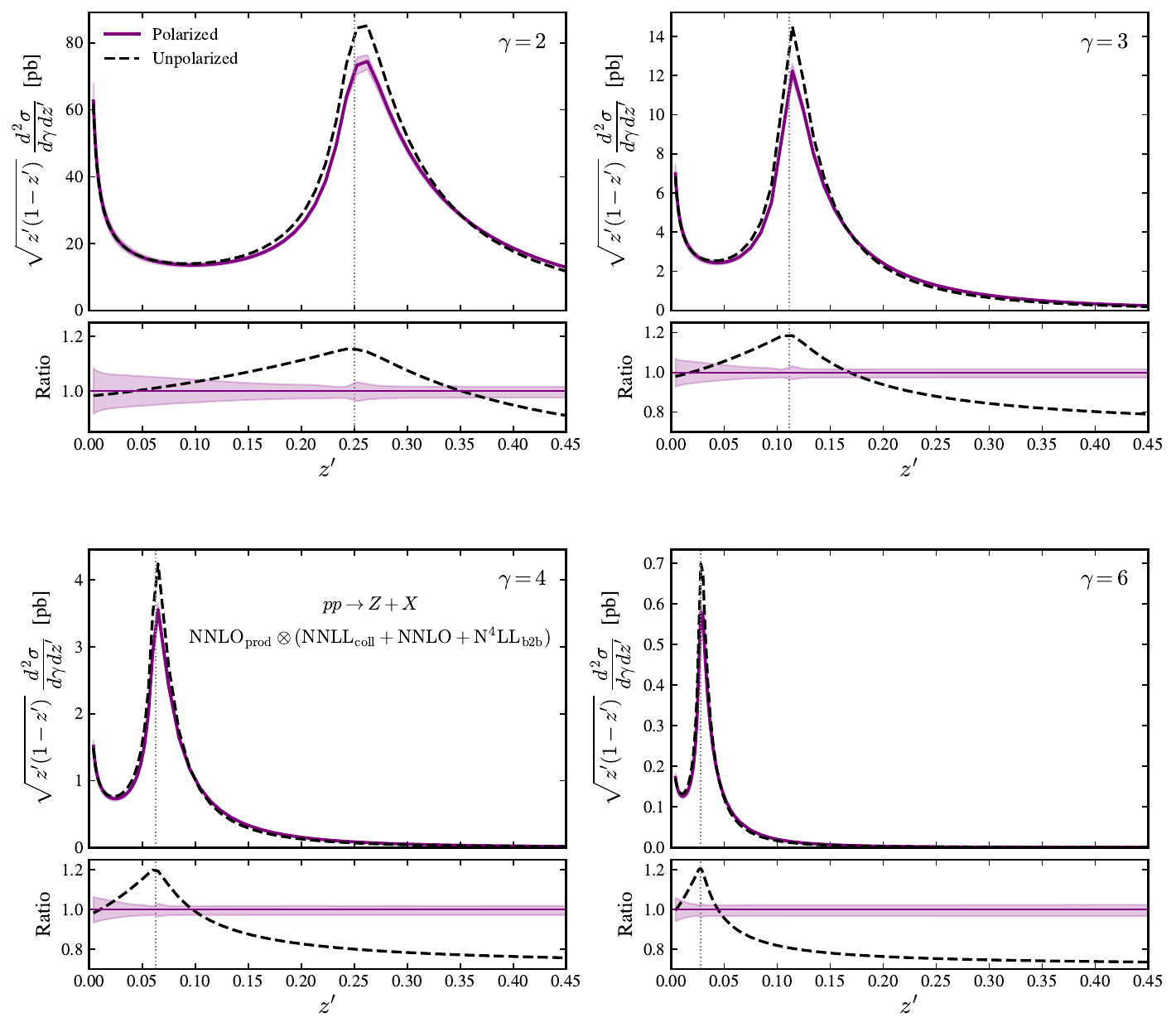}
\caption{\label{fig:fixedgamma}
Similar to Fig.~\ref{fig:money}, but for a fixed boost factor in $pp\to Z(\to q\bar q)+X$ at $\gamma=2$, $3$, $4$, and $6$, corresponding to $p_{T}\simeq158$, $258$, $353$, and $540$~GeV at $y_Z=0$, compared with the unpolarized reference $\rho_{00}=1/3$ at the same production rate (dashed). Dotted lines mark $z' = 1/\gamma^2$ and the naive unpolarized treatment overshoots by up to $\sim20\%$ at the decay peak.
}
\end{figure}

The EEC of the hadronic decay of $V$ in $pp \to V+X$ with $p_{T} > 100\,{\rm GeV}$ and $y_V=0$ is shown in Fig.~\ref{fig:money}. At fixed rapidity, the boost factor $\gamma$ and $p_{T}$ are mapped one to one, and the prediction is obtained by integrating Eq.~\eqref{eq:master} over the range of $\gamma$ corresponding to $p_{T} > 100\,{\rm GeV}$.
Compared with the naive unpolarized treatment, the correctly polarized spectrum is suppressed by up to ${\sim}10\%$ across the peak region, because high-$p_T$ production of $V+X$ is dominantly transverse. The $W^\pm$ ratio sits at smaller $z'$ than the $Z$ ratio because the same $p_T$ cut selects more boosted $W$ bosons.
Fig.~\ref{fig:fixedgamma} displays the same comparison but now at four fixed boost factors $\gamma$ (or the corresponding $p_{T}$), exhibiting the migration of the decay peak with $z' = 1/\gamma^2$ and the localization of the polarization effect at the peak. At fixed boost the deviation is even larger, reaching ${\sim}20\%$ at the peak, and it grows with the transverse dominance of the production density matrix.
The effect is not degenerate with a change of resummation or matching accuracy.
It has a characteristic shape, localized at the decay peak and reduced to a percent-level normalization shift in the collinear scaling region.
Effects of this size show that polarization must be treated correctly in any precision program for boosted $V$ bosons.

Conversely, the measured $z'$ spectrum provides a polarization analyzer complementary to measurements based on leptonic $W$ decays~\cite{CMS:2011kaj,ATLAS:2012au} and $Z$-decay angular distributions~\cite{CMS:2015cyj,ATLAS:2016rnf}, with different experimental systematics.

\subsection{\texorpdfstring{Polarized spectra in $e^+e^- \to W^+W^-$}{Polarized spectra in e+e- -> W+W-}}
\label{sec:eeprod}

\begin{figure}[t]
\centering
\includegraphics[width=\columnwidth]{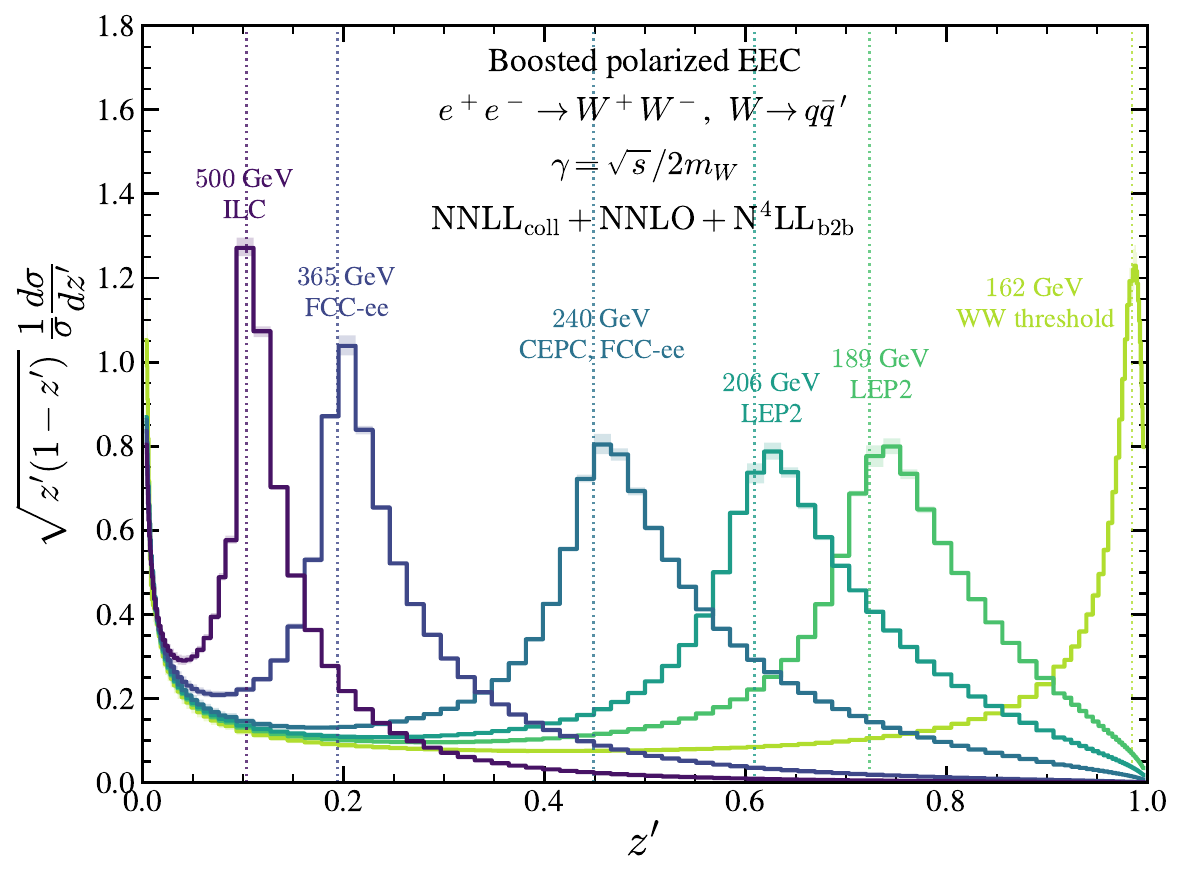}
\caption{ Normalized boosted EEC for the hadronically decaying $W$ (either $W^+$ or $W^-$)  in $e^+e^-\to W^+W^-$, at the center-of-mass energies displayed in the legend, with $\gamma=\sqrt{s}/(2m_W)$. Each curve is Eq.~\eqref{eq:rhocontractionboost} with the production value $\rho_{00}(\sqrt s)$, weighted by $\sqrt{z'(1-z')}$ for display. The production fractions are evaluated at leading electroweak order (tree level), using App.~\ref{app:eeWW}, for unpolarized beams and an inclusive leptonic decay, as in Tab.~\ref{tab:WWfractions}. The perturbative label refers to the decay correlator accuracy. }
\label{fig:eeWWboostEEC}
\end{figure}

Next, we consider boosted vector bosons produced at lepton colliders. For $e^+e^-\to W^+W^-$~\cite{Hagiwara:1986vm}, relevant for LEP2 legacy data and future lepton colliders, we consider semileptonic $W^+W^-$ events, with one $W$ decaying leptonically and the EEC measured on the hadronic decay of the other. Experimentally, such events provide a very clean $W^+W^-$ sample. For the predictions below, the leptonic decay serves only to identify the semileptonic channel and is integrated inclusively, with no fiducial cuts on its decay products. The boson energy is fixed by $\sqrt s$, so that up to initial-state radiation (ISR) effects the boost collapses to $\gamma = \sqrt{s}/(2 m_W)$, and the density matrix follows at leading order from the tree-level helicity amplitudes, whose angle-integrated squares we collect in App.~\ref{app:eeWW}. 
Tab.~\ref{tab:WWfractions} lists the diagonal entries $\rho_{\lambda\lambda}$ of the $W^-$ in the helicity frame at the energies shown in Fig.~\ref{fig:eeWWboostEEC}.
The polarization content is very different from high-$p_T$ hadroproduction. At $\sqrt{s}=162$~GeV, close to threshold, $\rho_{00}\simeq0.329$. It decreases to $0.0359$ at $500$~GeV, where the $W^-$ is predominantly $h=-1$ polarized, with $\rho_{--}\simeq0.926$.

\begin{table}[t]
\centering
\begin{tabular}{c|ccc}
\hline\hline
$\sqrt{s}$ & $\rho_{--}$ & $\rho_{00}$ & $\rho_{++}$ \\
\hline
162~GeV & 0.377 & 0.329 & 0.294 \\
189~GeV & 0.574 & 0.251 & 0.175 \\
206~GeV & 0.644 & 0.212 & 0.144 \\
240~GeV & 0.739 & 0.155 & 0.106 \\
365~GeV & 0.881 & 0.0644 & 0.0543 \\
500~GeV & 0.926 & 0.0359 & 0.0380 \\
\hline\hline
\end{tabular}
\caption{\label{tab:WWfractions}
Diagonal entries $\rho_{\lambda\lambda}$ of the helicity-frame density matrix of the $W^-$ in $e^+e^-\to W^+W^-$ with unpolarized beams. The subscripts $-$, $0$, and $+$ denote helicities $-1$, $0$, and $+1$, respectively. The energies match Fig.~\ref{fig:eeWWboostEEC}. The entries are calculated from leading-order electroweak (tree-level) production amplitudes using the analytic expressions and input parameters in App.~\ref{app:eeWW}. The $W^+$ fractions follow by interchanging the two transverse columns.
}
\end{table}

Fig.~\ref{fig:eeWWboostEEC} shows the resulting boosted EEC from the threshold region to $\sqrt s=500$~GeV.
The decay peak follows $z'\simeq1/\gamma^2$, from $z'\simeq1$ near threshold, where the $W$ is nearly at rest, to $z'\simeq0.1$ at $\sqrt s=500$~GeV, while the longitudinal fraction that enhances the peak decreases with energy, as in Tab.~\ref{tab:WWfractions}.

\section{Electroweak bosons without reconstruction}
\label{sec:bump}

The predictions above assume an identified boosted $V$, e.g.\ through explicit reconstruction of the resonance.
We now drop this assumption entirely.
Consider events with a leptonically decaying $Z$ ($Z\to\ell\ell$, serving only as a clean tag of the hard process) recoiling against a hadronic system. Energy correlators of the hadronic system recoiling against a $Z$ or a photon were recently proposed as a benchmark for precision QCD and computed at NNLL+NLO in Ref.~\cite{Generet:2026vto}.
The sample is then a mixture of $Z+$hadronic jet production, where the recoiling jet can be a QCD jet or a hadronically decaying $W$ or $Z$ from the $ZZ$ and $WZ$ processes.
On the recoiling hadronic jet, we can then measure the EEC inclusively, with no grooming, no mass window, and no tagging of any kind, normalized per selected event as
\begin{align}\label{eq:bumpobs}
\frac{1}{\sigma_{\rm sel}}&\frac{\dd\sigma}{\dd\ln\Delta R} \nonumber\\
&= \bigg\langle \sum_{i\neq j \in J} \frac{p_{T,i}\, p_{T,j}}{\big(p_{T}^{J}\big)^2}\;
\delta\big(\ln\Delta R - \ln\Delta R_{ij}\big) \bigg\rangle_{\rm events}\,,
\end{align}
with $\Delta R_{ij}^2 = \Delta\eta_{ij}^2 + \Delta\phi_{ij}^2$, in events selected only by the presence of a leptonic $Z$ and a recoiling anti-$k_T$ jet.

\begin{figure}[t]
\includegraphics[width=\columnwidth]{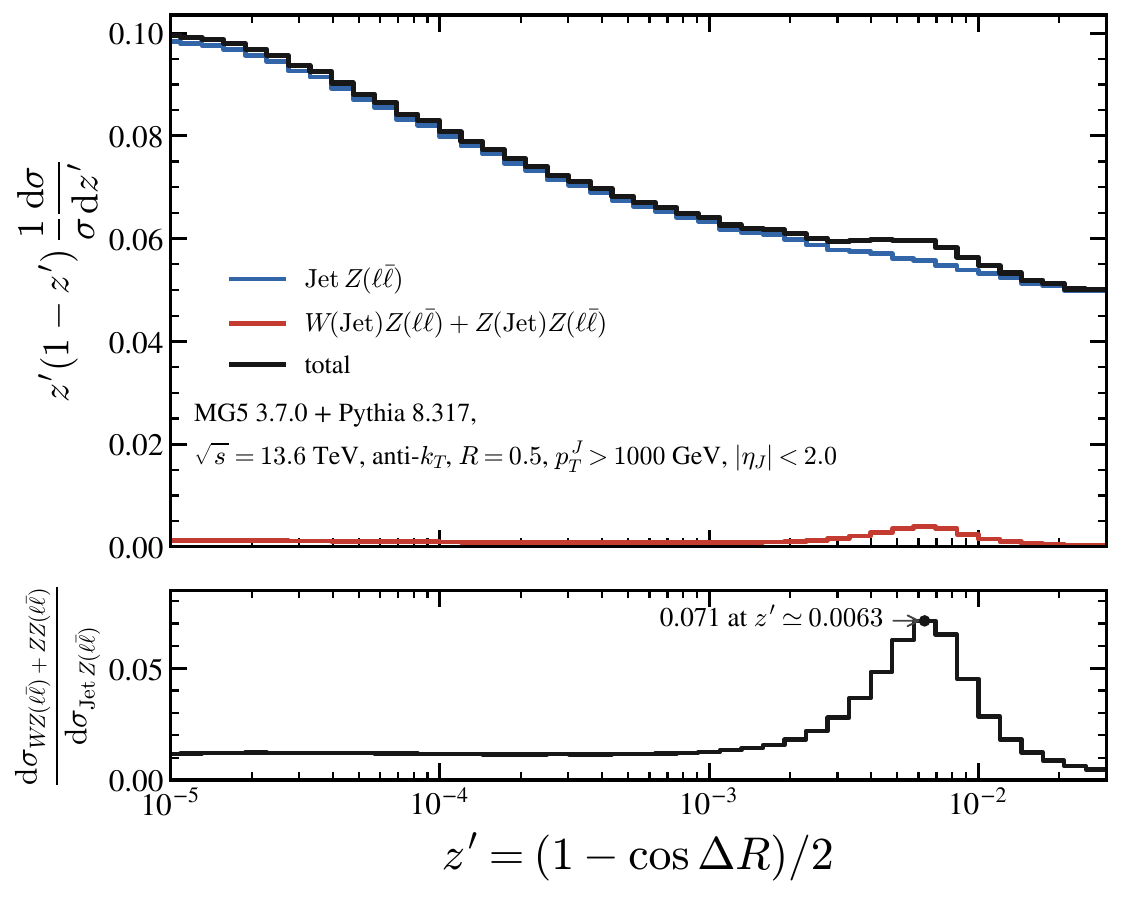}
\caption{\label{fig:bump}
EEC distribution on the hadronic anti-$k_T$ jet with $R=0.5$ and $p_T^J > 1$~TeV recoiling against a leptonic $Z(\to \ell \ell)$. The $Z(\ell\ell)+$jet QCD background alone (blue) is compared with the same sample including the diboson processes $WZ$ and $ZZ$ with hadronic $W/Z$ decays (black).
No reconstruction, grooming, or tagging is applied to the jet.
The lower panel shows the diboson-induced uplift, which peaks at $\approx 7\%$ near the characteristic angle $\Delta R \simeq 2 m_V/p_T^J$ even though strictly $W/Z$-originated jets constitute only $\approx 1\%$ of the sample.
}
\end{figure}

As an illustration, we carry out a Monte Carlo study by selecting the recoiling hadronic jet, clustered with the anti-$k_T$ algorithm with radius $R=0.5$ and required to have $p_T^J > 1$~TeV and $|\eta^J|<2.0$, as shown in Fig.~\ref{fig:bump}.
The QCD background exhibits the well-understood smooth scaling of the collinear EEC, $\sim 1/\Delta R$ up to logarithms~\cite{Dixon:2019uzg,Komiske:2022enw,Chen:2020vvp,Lee:2022uwt,Chen:2023zlx}.
For a two-body decay of a boson with transverse momentum $p_T^J$, the EEC distribution of the decay products exhibits a sharp peak near $\Delta R \simeq 2 m_V/p_T^J$.
This appears in Fig.~\ref{fig:bump} as a localized excess of around $7\%$, even though genuinely $W/Z$-initiated jets make up only around $1\%$ of the selected sample.
In effect, the $W$ and $Z$ are \emph{rediscovered} on top of the realistic QCD background, visible by eye as a resonance in angle rather than in mass, without ever being reconstructed.

The leptonic-$Z$ selection is chosen here for clarity rather than for reach.
It provides a clean demonstration of the electroweak bump on top of realistic QCD jets, but its limited rate makes it a target better suited to the HL-LHC than to existing datasets.
Fully inclusive dijet samples offer far larger statistics at the price of a smaller signal, with the SM $V$+jet contribution reduced to a percent-level feature relative to the dijet background.
Even so, the clean scaling of the QCD background and the strong correlations between angular bins preserve substantial discriminating power.
This is what allowed Ref.~\cite{Ricci:2026fkj}, recasting dijet events from the CMS $\alpha_s$ measurement~\cite{CMS:2024mlf}, to project competitive limits on light hadrophilic $Z'$ bosons \emph{without any tagging}, by treating the correlations between angular bins carefully. Such a careful study of the correlations between bins will also allow rediscovering the $W$ and $Z$ in dijet events.

These observations have immediate consequences for both searches and measurements.
For searches, our polarized, resummed templates upgrade such bump hunts into precision analyses.
The peak position calibrates the angular scale, making $m_V$ a standard candle inside jets~\cite{Holguin:2023bjf}, and the polarization-dependent peak shape provides exactly the spin discrimination of a hypothetical new resonance advocated in Ref.~\cite{Ricci:2026fkj}.
For measurements, a percent-level-understood QCD scaling baseline in QCD jets~\cite{Dixon:2019uzg,Komiske:2022enw,Chen:2020vvp,Lee:2022uwt,Chen:2023zlx} plus a first-principles polarized signal template turns the untagged $V$ bump into a laboratory for extracting electroweak boson rates, polarization fractions, and eventually couplings in hadronic final states.

\section{Conclusions and outlook}
\label{sec:conclusions}

We have initiated precision studies of energy correlators of boosted, polarized vector bosons.
In the rest frame of the boson, we decomposed the correlation matrix of the two-point energy correlator between helicity states into three real structure functions, $\Wcal^{(0,0)}$, $\Wcal^{(2,0)}$, and $\Wcal^{(2,2)}$, labeled by the rank $j$ of the spin operator and by its transverse spin $m$ about the bisector of the two detectors. Averaging over the azimuth about the boost axis removes the off-diagonal entries of the production density matrix, so that the opening-angle spectrum depends on the production only through its diagonal entries $\rho_{\lambda\lambda}$, and, since the two transverse helicities give the same boosted spectrum, only through the longitudinal fraction $\rho_{00}$. The off-diagonal entries become accessible once the orientation of the detectors relative to the production plane is measured. We illustrated this in App.~\ref{app:onepoint} with one-point correlators of the energy and charge flux, which together determine all eight real parameters of the density matrix, including the vector polarization that energy correlators cannot access. 

We computed these three structure functions region by region. In the back-to-back region, the contribution of the Collins functions vanishes for the energy-weighted correlator, so that the N$^4$LL resummation of $\Wcal^{(0,0)}$ carries over to all three. In the collinear region, the factorization involves hard functions of definite rank and jet functions of definite transverse spin. For $m=0$ the only new ingredient is the longitudinal coefficient function, which allows NNLL resummation, while the $m=2$ hard and jet functions, known only at their first order, drop out of the boosted spectrum at leading power in its collinear region. Away from these regions, $\Wcal^{(2,m)}$ follows from $\Wcal^{(0,0)}$ through the fixed-order ratios $a^{(2,m)}$, which we computed at NNLO. The resulting decay correlators combine NNLO fixed order with NNLL collinear and N$^4$LL back-to-back resummation, ${\rm NNLL_{coll}}+{\rm NNLO}+{\rm N^4LL_{b2b}}$.

Under the boost, each diagonal entry of the decay matrix becomes a sum of one-dimensional convolutions of the three structure functions with polarization-weighted boost kernels, which we derived in closed form in terms of complete elliptic integrals in App.~\ref{app:kernels}, together with their endpoint limits and the energy and momentum sum rules they preserve for each helicity.
The narrow-width factorization of Eq.~\eqref{eq:rhocontractionboost}, with the boosted spectra of Eqs.~\eqref{eq:W00} and~\eqref{eq:Wtransverse}, cleanly separates the polarized production density matrices and the precision rest-frame correlators, each of which can be improved independently.
The polarization dependence is concentrated at the decay peak $z'\simeq1/\gamma^2$, which is fed by decays perpendicular to the boost axis, the preferred orientation of a longitudinal boson, so that at $\gamma=4$ the peak height differs by about $70\%$ between longitudinal and transverse bosons. For the production, we computed the polarized density matrices of $pp\to W^\pm,Z+X$ at NNLO in QCD and those of $e^+e^-\to W^+W^-$ at leading order, for which we gave closed forms. For high-$p_T$ $W$ and $Z$ bosons in $pp$ collisions, polarization effects enter boosted-$V$ spectra at the $10$--$20\%$ level with a distinctive shape, while in $e^+e^-\to W^+W^-$ they are negligible near threshold, where $\rho_{00}\simeq1/3$, and grow with energy as the $W$ bosons become increasingly transverse. These spectra turn the hadronic decays of boosted $W$ and $Z$ bosons into polarimeters, complementary to the leptonic decays on which polarization measurements have relied so far.

Finally, we showed that hadronic $W$ and $Z$ decays are visible without any reconstruction. In untagged jets recoiling against a leptonic $Z$, they appear in the EEC as a localized bump of about $7\%$ at $\Delta R\simeq2m_V/p_T$ on top of the smooth QCD scaling, although they make up only about $1\%$ of the jets.

Many directions follow. New-physics effects that change the production rates of the different helicities, studied in Ref.~\cite{Ricci:2022htc}, can now be revisited with precision predictions.
Higher-point correlators, whose rest-frame technology exists for scalar decays~\cite{Gao:2024wcg} and whose spinning structures are classified in Ref.~\cite{Riembau:2025isw}, probe the polarized decay in richer detail. Generalized detectors~\cite{Gonzalez:2026aha}, which measure powers of the energy on subsets of hadrons, offer further handles, since higher energy weights can enhance the signals from hadronic decays of electroweak bosons or new-physics resonances relative to QCD jet backgrounds.
Boosted top decays inherit a polarized $W$, providing a standard candle inside top jets~\cite{Holguin:2022epo,Holguin:2023bjf}.
The relevant datasets, from high-$p_T$ $V$+jets to diboson and inclusive jet samples, already exist, and the framework developed here provides the theoretical precision needed for their analysis.

\begin{acknowledgments}
We thank Hao Chen, Lance Dixon, Terry Generet, Phil Harris, Ian Moult, Lorenzo Ricci, and Simon Rothman for helpful discussions. We thank Lance Dixon and Ian Moult for providing the \textsc{CoLoRFulNNLO} data for EEC at NNLO. AG is supported by the U.S. Department of Energy under contract DE-AC02-76SF00515. KL is supported by the U.S. Department of Energy under contract DE-AC02-06CH11357. XYZ is supported by the MIT Pappalardo Fellowship. 
This work was also supported in part by the Aspen Center for Physics (NSF grant PHY-2210452) and the Erwin-Schrödinger International Institute for Mathematics and Physics. The computations in this paper were run on the FASRC Cannon cluster supported by the FAS Division of Science Research Computing Group at Harvard University, the subMIT facility at the MIT Physics Department~\cite{Acosta:2025kgk}, and the SLAC Shared Science Data Facility (S3DF) at SLAC National Accelerator Laboratory.

\end{acknowledgments}

\appendix

\section{Closed forms of the weighted boost kernels}
\label{app:kernels}

Here we give closed forms for the three weighted kernels entering the
diagonal helicity spectra in Eq.~\eqref{eq:W00}.  The formulas below apply for
$\gamma>1$ and $0<z,z'<1$; the distributional endpoint limits are given in Eqs.~\eqref{eq:appz0} and~\eqref{eq:appz1}. 

To compute the integral in Eq.~\eqref{eq:Kw} with the three weights of Eq.~\eqref{eq:weights}, we first carry out the $\varphi$ integral to remove the delta function.
We set $u=\cos\vartheta$, which runs over $[-1,1]$, and write $\lambda_1\lambda_2$ as
\begin{align}\label{eq:lam1lam2}
\lambda_1\lambda_2
&=A(u)^2-(\gamma^2-1)z(1-u^2)\cos^2\varphi\,,
\end{align}
where
\begin{align}\label{eq:Au}
    A(u)&\equiv\gamma+\sqrt{(\gamma^2-1)(1-z)}\,u
    \,.
\end{align}
The delta function fixes 
\begin{align}\label{eq:Xu}
\cos^2\varphi=X(u)\equiv\frac{A(u)^2-z/z'}{(\gamma^2-1)z(1-u^2)}\,,
\end{align}
so real solutions require
$0\leq X(u)\leq1$.  Summing over these solutions in the $\varphi$ integral
leaves the Jacobian factor $1/\sqrt{Q(u)}$ where
\begin{align}
Q(u)&\equiv\bigl(A(u)^2-z/z'\bigr)\nonumber\\
&\quad\times\Bigl[(\gamma^2-1)z(1-u^2)-A(u)^2+z/z'\Bigr]\,.
\end{align}
$Q(u)$ vanishes exactly where $\cos^2\varphi=X(u)$ reaches an end of its allowed range $0\leq X(u)\leq1$. These points bound the $u$ integration, and the Jacobian $1/\sqrt{Q(u)}$ has integrable singularities there.
 The four roots of $Q(u)$ are $r_{1,2}$, where $X(u)=0$, and $r_{3,4}$, where $X(u)=1$,
\begin{align}\label{eq:apprawroots}
r_{1,2}&=-\frac{\gamma\pm\sqrt{z/z'}}{\sqrt{(\gamma^2-1)(1-z)}}\,,\nonumber\\
r_{3,4}&=-\frac{\gamma\sqrt{1-z}\pm\sqrt{z/z'-z}}{\sqrt{\gamma^2-1}}\,.
\end{align}
We label these roots as $u_1,\cdots ,u_4$ in increasing order, $u_1<u_2<u_3<u_4$. In terms of the
unsorted roots in Eq.~\eqref{eq:apprawroots}, the ordering on the two branches is
\begin{align}\label{eq:approotorder}
(u_1,u_2,u_3,u_4)=
\begin{cases}
(r_1,r_3,r_2,r_4)\,,&0<\xi(z,z')<1\,,\\
(r_1,r_2,r_3,r_4)\,,&\xi(z,z')>1\,,
\end{cases}
\end{align}
where 
\begin{align}\label{eq:appmdef}
\xi(z,z')=
\frac{1}{2}+\frac{z+z'+zz'(\gamma^2-1)-2}{4\sqrt{(1-z)(1-z')}}\,.
\end{align}
The two branches are the regions $\xi<1$ and $\xi>1$ of the $(z,z')$ plane.
For either ordering, $0\leq X(u)\leq1$ within the physical range $|u|\leq1$ selects the domain
$u_3\leq u\leq u_4$. At $\xi=1$ the roots $r_2$ and $r_3$ coincide at $u=-1$ and exchange their places in the ordering.

With Eq.~\eqref{eq:Xu}, the three weights in Eq.~\eqref{eq:weights}
reduce to
\begin{align}\label{eq:appweights1d}
w(u)\in\bigl\{1,\ 3u^2-1,\
(1-u^2)\bigl[2X(u)-1\bigr]\bigr\}\,.
\end{align}
Performing the $\varphi$ integral therefore gives
\begin{align}\label{eq:appweightedkernel1d}
\Kcal_\gamma^{[w]}(z,z')
&=\frac{z^2}{2\pi z'^3\gamma^2}
\int_{u_3}^{u_4}\frac{\dd u}{\sqrt{Q(u)}}\,w(u)\,.
\end{align}

Now we introduce the elliptic modulus $k$ and the characteristic $\nu$ as
\begin{align}\label{eq:appellipticdata}
k^2&=\frac{(u_1-u_2)(u_3-u_4)}{(u_1-u_3)(u_2-u_4)}\,,&
\nu&=-\frac{u_3-u_4}{u_1-u_3}\,,
\end{align}
which are the arguments of the complete elliptic integrals of the three kinds defined in Eq.~\eqref{eq:appellipticdefs} below.
With the ordering in Eq.~\eqref{eq:approotorder},
$k^2=\xi(z,z')$ for $0<\xi(z,z')<1$, while
$k^2=1/\xi(z,z')$ for $\xi(z,z')>1$.
The three weighted kernels can then be expressed in terms of complete elliptic integrals,
\begin{align}\label{eq:appKelliptic1}
&\Kcal_\gamma^{[1]}(z,z')
=\frac{z^2K(k)}{\pi z'^3\gamma^2(\gamma^2-1)
\sqrt{(1-z)(u_1-u_3)(u_2-u_4)}}\,,
\\[0.2cm]
\label{eq:appKellipticP2}
&\Kcal_\gamma^{[3\cos^2\vartheta-1]}(z,z')=\nonumber\\&\quad
\frac{z^2}{\pi z'^3\gamma^2(\gamma^2-1)
\sqrt{(1-z)(u_1-u_3)(u_2-u_4)}}\nonumber\\&\qquad
\times\Big\{(3u_1^2-1)K(k)
+6u_1(u_4-u_1)\Pi(\nu,k)\nonumber\\&\qquad\qquad
+3(u_4-u_1)^2\Pi_2(\nu,k)\Big\}\,,
\\[0.3cm]
\label{eq:appKelliptic2}
&\Kcal_\gamma^{[\sin^2\vartheta\cos2\varphi]}(z,z')=\nonumber\\&\quad
\frac{z}{\pi z'^3\gamma^2(\gamma^2-1)^2
\sqrt{(1-z)(u_1-u_3)(u_2-u_4)}}\nonumber\\
&\times\Big\{\big[(\gamma^2-1)(2-z)u_1^2
+4\gamma\sqrt{(\gamma^2-1)(1-z)}\,u_1\nonumber\\
&\qquad\qquad\qquad+2\gamma^2-2z/z'-(\gamma^2-1)z\big]K(k)\nonumber\\
&\qquad+2(u_4-u_1)\Big[(\gamma^2-1)(2-z)u_1\nonumber\\
&\qquad\qquad\qquad+2\gamma\sqrt{(\gamma^2-1)(1-z)}\Big]\Pi(\nu,k)\nonumber\\
&\qquad+(\gamma^2-1)(2-z)(u_4-u_1)^2
\Pi_2(\nu,k)\Big\}\,.
\end{align}
Note that Eq.~\eqref{eq:appKelliptic1} is algebraically identical to Eq.~\eqref{eq:Kscalar} on both
branches.

Here, the complete
elliptic integrals are
\begin{align}\label{eq:appellipticdefs}
K(k)&=\int_0^{\pi/2}\frac{\dd\alpha}{\sqrt{1-k^2\sin^2\alpha}}\,,\nonumber\\
E(k)&=\int_0^{\pi/2}\dd\alpha\,
\sqrt{1-k^2\sin^2\alpha}\,,\nonumber\\
\Pi(\nu,k)&=\int_0^{\pi/2}\frac{\dd\alpha}{(1-\nu\sin^2\alpha)\sqrt{1-k^2\sin^2\alpha}}\,.
\end{align}
And the second-order generalized complete elliptic integral of the third kind we need is
\begin{align}\label{eq:appPi2def}
\Pi_2(\nu,k)
&\equiv\int_0^{\pi/2}\frac{\dd\alpha}{(1-\nu\sin^2\alpha)^2\sqrt{1-k^2\sin^2\alpha}}\nonumber\\
&=\partial_\nu\bigl[\nu\Pi(\nu,k)\bigr]\nonumber\\
&=\frac{\nu E(k)+(k^2-\nu)K(k)}{2(k^2-\nu)(\nu-1)}\nonumber\\
&\quad+\frac{2\nu k^2-3k^2-\nu^2+2\nu}{2(k^2-\nu)(\nu-1)}\Pi(\nu,k)\,.
\end{align}

Solving $0<\xi(z,z')<1$ and $\xi(z,z')>1$ at fixed $\gamma>1$ and
$0<z<1$ determines the ranges of $z'$ for the two branches.  We briefly
repeat the analysis of Ref.~\cite{Gao:2026xuq} below.  Define
\begin{align}
z'_\pm=\frac{z}{\bigl(\gamma\mp\sqrt{(\gamma^2-1)(1-z)}\bigr)^2}\,.
\end{align}
The kinematically allowed intervals are
\begin{align}\label{eq:appsupport}
z'_-\leq z'\leq z'_+&\,,&&0<z\leq\gamma^{-2}\,,\nonumber\\
z'_-\leq z'\leq1&\,,&&\gamma^{-2}<z<1\,.
\end{align}
The kernels vanish outside the intervals in Eq.~\eqref{eq:appsupport}.

The two cases correspond to different branch structures:
\begin{itemize}
\item For $0<z\leq\gamma^{-2}$, $z'_+$ is the upper kinematic boundary.
Only the $0\leq \xi<1$ branch contributes, with $k^2=\xi$.
\item For $\gamma^{-2}<z<1$, the allowed interval extends to $z'=1$,
and $z'_+$ is an internal branch point.  The region
$z'_-\leq z'<z'_+$ corresponds to $0\leq \xi<1$, with $k^2=\xi$,
whereas $z'_+<z'<1$ corresponds to $\xi>1$, with $k^2=1/\xi$.
At $z'=z'_+$, where $\xi=1$, the kernels in
Eqs.~\eqref{eq:appKelliptic1} and~\eqref{eq:appKellipticP2} have an
integrable logarithmic singularity. The kernel in
Eq.~\eqref{eq:appKelliptic2} remains finite there because its weight
$w=\sin^2\vartheta\cos2\varphi$ vanishes at $u=-1$ where $r_2$ and $r_3$
coalesce, removing the logarithmic singularity.
\end{itemize}

The boost kernels take a particularly simple form in certain kinematic limits. In the zero-boost limit,
\begin{align}\label{eq:appzeroboost}
\Kcal_{\gamma=1}^{[1]}(z,z')&=\delta(z'-z)\,,\nonumber\\
\Kcal_{\gamma=1}^{[3\cos^2\vartheta-1]}(z,z')&=0\,,\nonumber\\
\Kcal_{\gamma=1}^{[\sin^2\vartheta\cos2\varphi]}(z,z')&=0\,.
\end{align}
At the collinear endpoint one finds
\begin{align}\label{eq:appz0}
\Kcal_\gamma^{[1]}(0,z')
&=\frac{4\gamma^2-1}{3\gamma^2}\,\delta(z')\,,\nonumber\\
\Kcal_\gamma^{[3\cos^2\vartheta-1]}(0,z')
&=\frac{4(\gamma^2-1)}{15\gamma^2}\,\delta(z')\,,\nonumber\\
\Kcal_\gamma^{[\sin^2\vartheta\cos2\varphi]}(0,z')&=0\,.
\end{align}
Finally, at the back-to-back endpoint we have
\begin{align}\label{eq:appz1}
\Kcal_\gamma^{[1]}(1,z')
&=\frac{\Theta(\gamma^2z'-1)}{2\gamma^2\sqrt{\gamma^2-1}\,z'^{5/2}
\sqrt{\gamma^2z'-1}}\,,
\nonumber\\
\Kcal_\gamma^{[3\cos^2\vartheta-1]}(1,z')
&=\frac{1}{2}\left(1-3\frac{\gamma^2z'-1}{(\gamma^2-1)z'}\right)
\Kcal_\gamma^{[1]}(1,z')\,,\nonumber\\
\Kcal_\gamma^{[\sin^2\vartheta\cos2\varphi]}(1,z')
&=-\frac{1}{2}\left(1-3\frac{\gamma^2z'-1}{(\gamma^2-1)z'}\right)
\Kcal_\gamma^{[1]}(1,z')\,.
\end{align}

\subsection{Kernel moments and EEC sum rules}

The zeroth and first moments in $z'$ provide useful checks on the EEC sum rules
after a boost.  They are most easily obtained before evaluating the angular
integrals in Eq.~\eqref{eq:Kw}.  Performing the $z'$ integral against the
delta function gives
\begin{align}
\int_0^1\!\dd z'\,\Kcal_\gamma^{[w]}(z,z')
&=\frac{1}{\gamma^2}\int\!\frac{\dd^2\Omega}{4\pi}\,
\lambda_1\lambda_2\,w\,,\nonumber\\
\int_0^1\!\dd z'\,z'\Kcal_\gamma^{[w]}(z,z')
&=\frac{z}{\gamma^2}\int\!\frac{\dd^2\Omega}{4\pi}\,w\,.
\end{align}
For the three weights in Eq.~\eqref{eq:weights}, the angular averages yield
\begin{align}\label{eq:appkernelmoments}
\int_0^1\!\dd z'\,\Kcal_\gamma^{[1]}(z,z')
&=1-\frac{2(\gamma^2-1)}{3\gamma^2}
\left(z-\frac{1}{2}\right)\,,\nonumber\\
\int_0^1\!\dd z'\,
\Kcal_\gamma^{[3\cos^2\vartheta-1]}(z,z')
&=\frac{2(\gamma^2-1)}{15\gamma^2}(2-z)\,,\nonumber\\
\int_0^1\!\dd z'\,
\Kcal_\gamma^{[\sin^2\vartheta\cos2\varphi]}(z,z')
&=-\frac{2(\gamma^2-1)}{15\gamma^2}z\,,
\end{align}
and
\begin{align}\label{eq:appkernelzpmoments}
\int_0^1\!\dd z'\,z'\Kcal_\gamma^{[1]}(z,z')
&=\frac{z}{\gamma^2}\,,\nonumber\\
\int_0^1\!\dd z'\,z'
\Kcal_\gamma^{[3\cos^2\vartheta-1]}(z,z')
&=0\,,\nonumber\\
\int_0^1\!\dd z'\,z'
\Kcal_\gamma^{[\sin^2\vartheta\cos2\varphi]}(z,z')
&=0\,.
\end{align}
These identities include the distributional endpoint limits in
Eqs.~\eqref{eq:appz0} and~\eqref{eq:appz1}.

We now use these moments to show that the boosted spectra obey the energy and momentum sum rules separately for each helicity. In the rest frame, the orientation average of Eq.~\eqref{eq:orientationaverage} reduces every diagonal entry to $\Wcal^{(0,0)}$, so the rest-frame sum rules are carried by $\Wcal^{(0,0)}$ alone. The structure function $\Wcal^{(0,0)}$ obeys the energy and momentum sum
rules
  \begin{align}\label{eq:apprestsumrules}
\int_0^1\!\dd z\,\Wcal^{(0,0)}(z)
=2\int_0^1\!\dd z\,z\Wcal^{(0,0)}(z)=1\,.
\end{align} 
After the boost, the diagonal spectra of Eqs.~\eqref{eq:W00} and~\eqref{eq:Wtransverse} also involve $\Wcal^{(2,0)}$ and $\Wcal^{(2,2)}$, through weighted kernels whose zeroth moments in Eq.~\eqref{eq:appkernelmoments} do not vanish. It suffices to consider the longitudinal spectrum, since Eqs.~\eqref{eq:W00} and~\eqref{eq:Wtransverse} give $\Wcal^{\rm boost}_{\pm\pm}=\tfrac12\big(3\,\Kcal_\gamma^{[1]}\otimes\Wcal^{(0,0)}-\Wcal^{\rm boost}_{00}\big)$, where $\otimes$ denotes the $z$ integral of Eq.~\eqref{eq:W00}.
Using Eqs.~\eqref{eq:W00} and~\eqref{eq:appkernelmoments}, together with Eq.~\eqref{eq:apprestsumrules} for the $\Wcal^{(0,0)}$ term, the longitudinal
boosted spectrum satisfies
  \begin{multline}\label{eq:appW00norm}
\int_0^1\!\dd z'\,\Wcal^{\rm boost}_{00}(z';\gamma)
=1+\frac{2(\gamma^2-1)}{15\gamma^2}\int_0^1\!\dd z\\
\times\Big[(2-z)\Wcal^{(2,0)}(z)
-\tfrac32 z\Wcal^{(2,2)}(z)\Big]\,.
\end{multline} 
Consequently, the energy sum rule $\int_0^1\dd z'\,\Wcal^{\rm boost}_{00}=1$ is preserved
provided the polarized rest-frame functions obey\footnote{Eq.~\eqref{eq:apppolarizedsumrule} follows from Eq.~(80) of Ref.~\cite{Riembau:2025isw}. Specializing it to the EEC gives $\int\!\dd\Omega_1 n_1^r H_{\mathcal{EE}}^{ab}=0$, where $H_{\mathcal{EE}}^{ab}$ is the EEC tensor defined in Eq.~(32) there, and projecting it with $\delta_{ra}n_{2b}$ yields Eq.~\eqref{eq:apppolarizedsumrule}.}
  \begin{align}\label{eq:apppolarizedsumrule}
\int_0^1\!\dd z\,\Big[(2-z)\Wcal^{(2,0)}(z)
-\tfrac32 z\Wcal^{(2,2)}(z)\Big]=0\,.
\end{align} 
The transverse spectra then integrate to $\tfrac12(3-1)=1$ as well.
The momentum sum rule requires no additional polarized constraint because
both weighted moments in Eq.~\eqref{eq:appkernelzpmoments} vanish, so only the $\Wcal^{(0,0)}$ term contributes. It enters all three diagonal spectra with unit coefficient, and Eqs.~\eqref{eq:appkernelzpmoments} and~\eqref{eq:apprestsumrules} give $2\gamma^2\int_0^1\dd z'\,z'\,\Kcal_\gamma^{[1]}\otimes\Wcal^{(0,0)}=2\int_0^1\dd z\,z\,\Wcal^{(0,0)}=1$. The factor $2\gamma^2$ appears because the first moment in $z'$ is fixed by the invariant mass of the decay products, $\int_0^1\dd z'\,z'\,\Wcal^{\rm boost}_{\lambda\lambda}=m_V^2/(2E_V^2)=1/(2\gamma^2)$. Thus
  \begin{align}\label{eq:appboostedsumrules}
\int_0^1\!\dd z'\,\Wcal^{\rm boost}_{\lambda\lambda}(z';\gamma)
&=2\gamma^2\int_0^1\!\dd z'\,z'
\Wcal^{\rm boost}_{\lambda\lambda}(z';\gamma)=1\,,
\end{align} 
for $\lambda=0,\pm$.

\section{Polarized \texorpdfstring{$W^\pm$}{W} production in \texorpdfstring{$pp$}{pp} collisions}
\label{app:Wprod}

In the main text, we present the production rate and boosted EEC distributions for the $Z$ boson. Here, we also provide the results for $W^{\pm}$ bosons. 

\begin{figure}[t]
\centering
\includegraphics[width=\columnwidth]{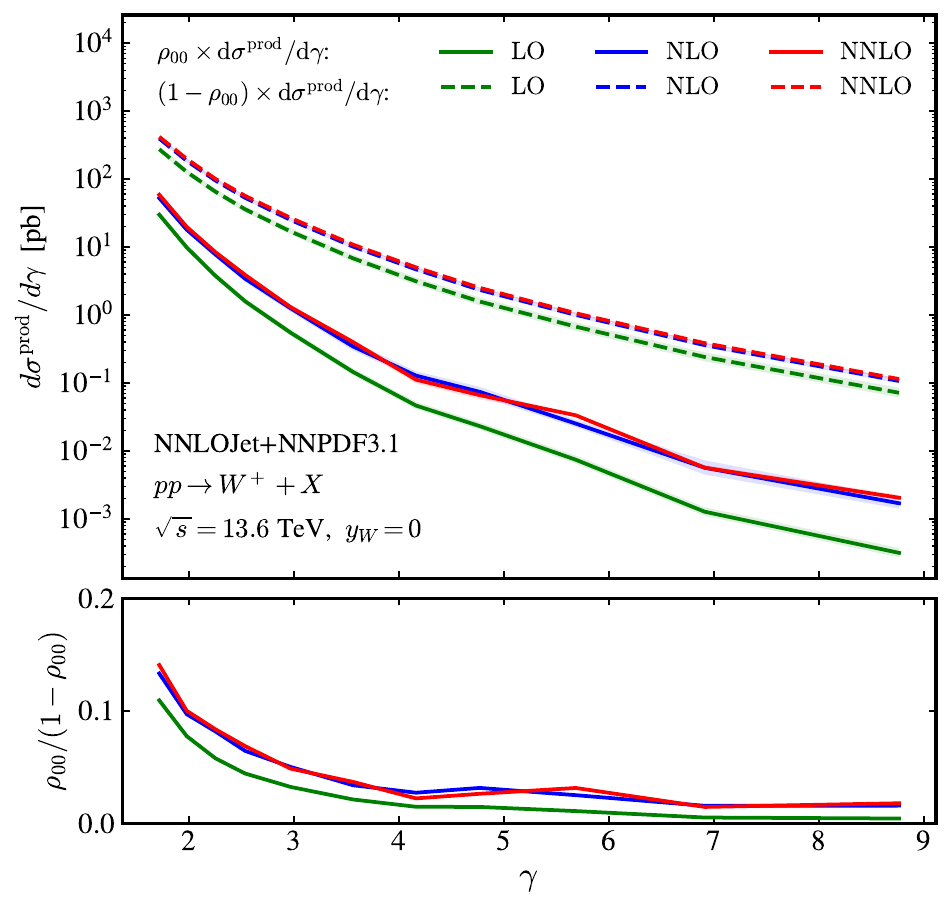}
\includegraphics[width=\columnwidth]{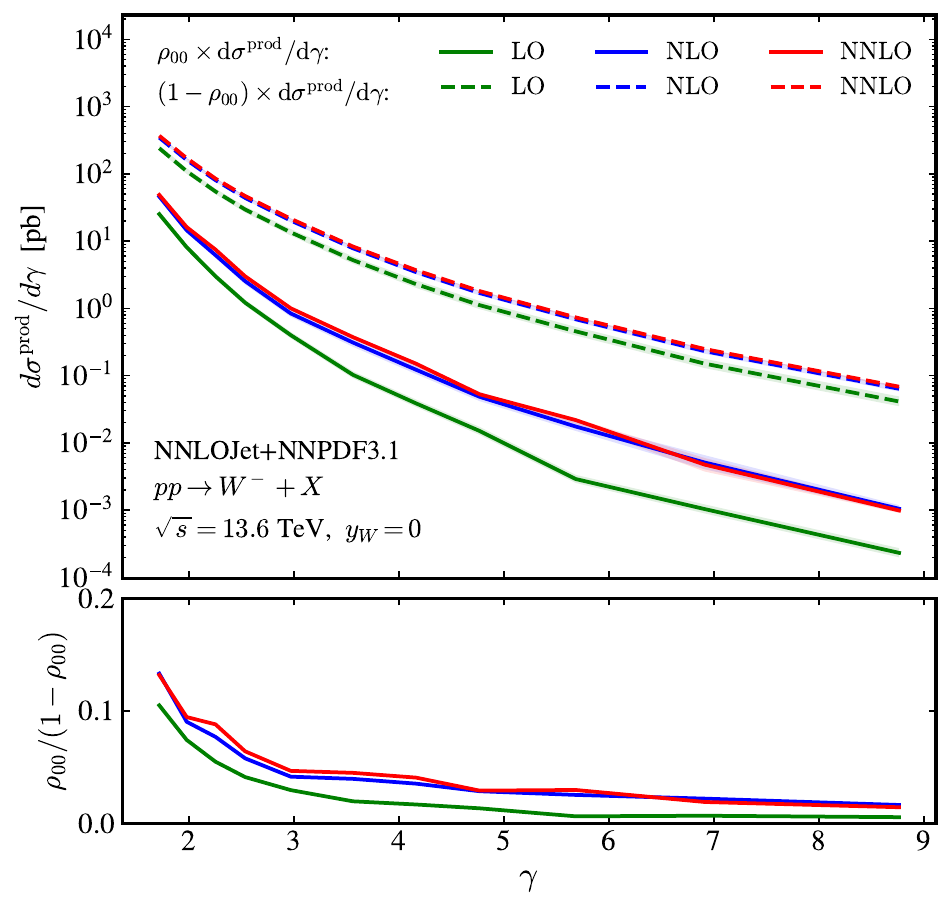}
\caption{ Longitudinal and total transverse production rates for $pp\to W^++X$ (upper plot) and $pp\to W^-+X$ (lower plot), with the collider settings indicated in the panels. The solid and dashed curves denote the longitudinal rate $\dd\sigma^{\rm prod}_{00}/\dd\gamma$ and the transverse rate $\sum_\pm\dd\sigma^{\rm prod}_{\pm\pm}/\dd\gamma$, respectively. Each ratio panel shows $\rho_{00}/(1-\rho_{00})=\sigma_{00}/(\sigma_{++}+\sigma_{--})$. The line-style convention and helicity frame are the same as in Fig.~\ref{fig:LTproduction}.}
  \label{fig:LTproductionW}
 \end{figure}

Fig.~\ref{fig:LTproductionW} shows the longitudinal and total transverse production rates and the ratio $\rho_{00}/(1-\rho_{00})$ for $pp\to W^++X$ and $pp\to W^-+X$ at $\sqrt s=13.6$~TeV and $y_W=0$, using \textsc{Nnlojet}. Similar to Fig.~\ref{fig:LTproduction}, we restore stable $W^{\pm}$ bosons and apply the same $p_T>100$~GeV cut.

\begin{figure}[t]
\includegraphics[width=\columnwidth]{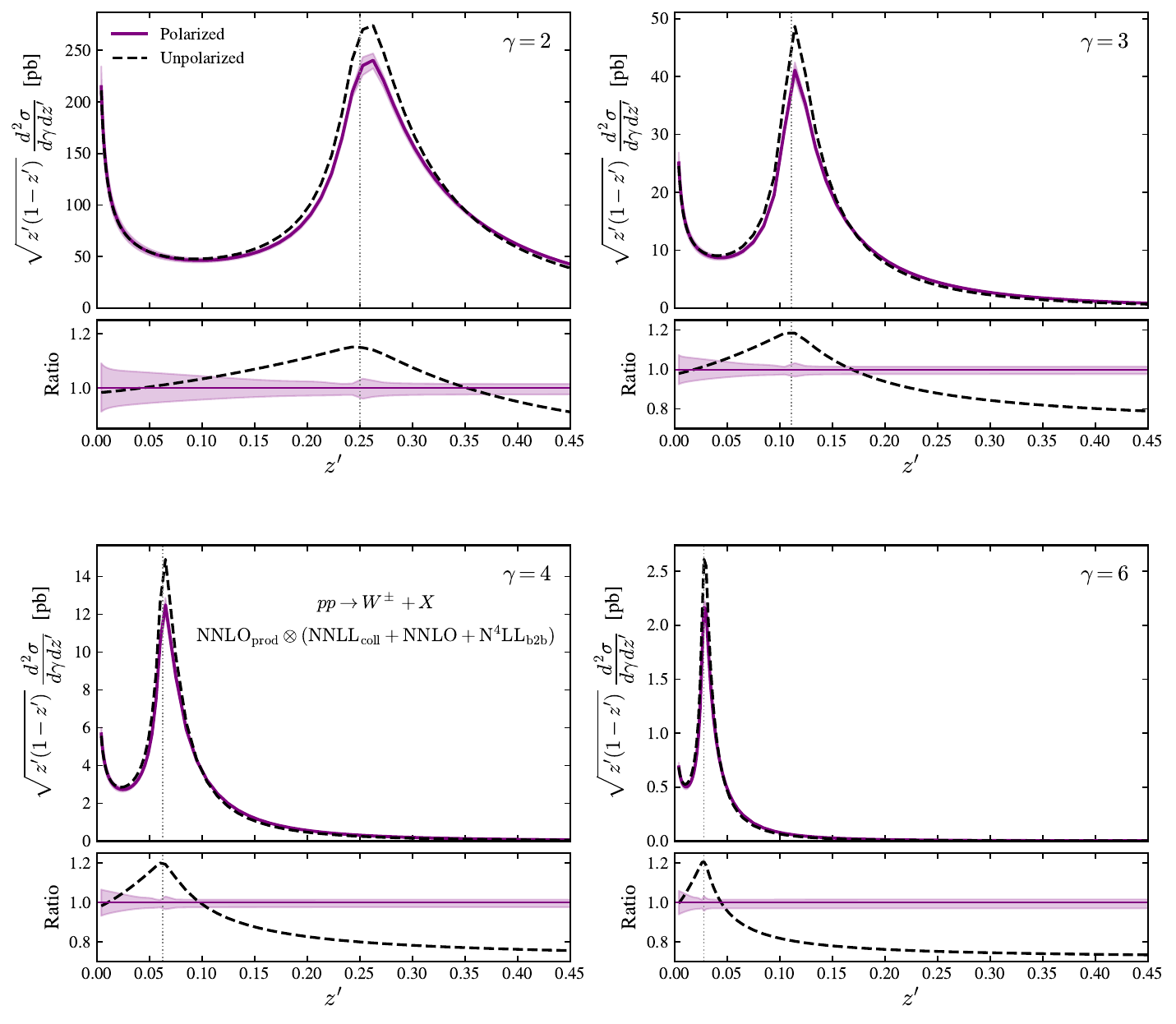}
\caption{\label{fig:fixedgammaW}
Similar to Fig.~\ref{fig:fixedgamma}, but for $pp\to W^{\pm}+X$. We choose $\gamma=2$, $3$, $4$, and $6$, corresponding to $p_{T}\simeq 139$, $227$, $311$, and $476$~GeV at $y_W=0$.
}
\end{figure}

Fig.~\ref{fig:fixedgammaW} gives the boosted EEC distributions for $pp\to W^\pm +X$ at fixed boost factors $\gamma=2,3,4,6$. Similar to Fig.~\ref{fig:fixedgamma}, we observe that the peak location moves together with the values of boost $\gamma$, and that the deviations from the naive unpolarized sum reach $20\%$ at the peak.

\section{Polarized $W$ production in $e^+e^-\to W^+W^-$}
\label{app:eeWW}

For convenience, we collect closed forms for the diagonal helicity fractions at leading order in the electroweak couplings, obtained by integrating the squares of the Standard Model tree-level (Born) helicity amplitudes of Ref.~\cite{Hagiwara:1986vm} over the production angle. At this order, $e^+e^-\to W^+W^-$ is a purely electroweak process, mediated by $s$-channel photon and $Z$ exchange and $t$-channel neutrino exchange. The amplitudes are $\mathcal O(e^2)$ and the polarized production cross sections are $\mathcal O(\alpha^2)$, with $\alpha=e^2/(4\pi)$; the fractions are ratios of these Born cross sections. Electroweak radiative corrections are not included. Since neither the beams nor the $W$ bosons carry color, QCD does not correct the Born process and enters the production only through quark loops in the electroweak corrections. This production accuracy is separate from the higher-order QCD accuracy of the hadronic decay correlators, which factorize from the production in the narrow-width approximation and which we compute at ${\rm NNLL_{coll}} + {\rm N^4LL_{b2b}} + {\rm NNLO}$ accuracy. The boosted EEC then follows from Eq.~\eqref{eq:rhocontractionboost}, in which the production fractions multiply the boosted longitudinal and transverse decay spectra.
We assume unpolarized beams, massless electrons, on-shell $W$ bosons, zero widths, no initial-state radiation, and no angular cuts. Each helicity is defined along the corresponding $W$ momentum in the $e^+e^-$ center-of-mass frame.

The single-$W^-$ fractions are
\begin{equation}\label{eq:eeWWfractions}
\rho^{W^-}_{\lambda\lambda}(s)=
\frac{\displaystyle\sum_{h,\bar\lambda}\int_{-1}^{1}\dd \cos\Theta\,
 |\mathcal M_{h;\lambda\bar\lambda}(s,\Theta)|^2}
 {\displaystyle\sum_{h,\mu,\bar\lambda}\int_{-1}^{1}\dd \cos\Theta\,
 |\mathcal M_{h;\mu\bar\lambda}(s,\Theta)|^2}\,.
\end{equation}
Here $\Theta$ is the angle between the incoming electron and the $W^-$, $h$ labels the helicity configurations of the incoming $e^+e^-$ pair, $\lambda$ and $\mu$ denote $W^-$ helicities, and $\bar\lambda$ denotes the $W^+$ helicity. The fractions $\rho^{W^+}_{\lambda\lambda}$ of the $W^+$ are defined in the same way, with the roles of the two bosons interchanged, and follow from Eq.~\eqref{eq:eeWWcharge} below.
An inclusive leptonic partner decay gives this spin trace in the narrow-width approximation. Cuts on the leptonic decay products would instead require an acceptance-weighted density matrix. We impose no such cuts, which corresponds to full acceptance for the leptonic decay, as in Sec.~\ref{sec:eeprod}.

For $s>4m_W^2$, the fractions depend on $x=\beta^2=1-4m_W^2/s$, with $\beta$ the $W$ velocity in the center-of-mass frame, and on
\begin{align}\label{eq:eeWWvariables}
a&=\frac{(1-2\sin^2\theta_W)\,s\,m_Z^2}{4m_W^2(s-m_Z^2)}\,,&
b&=\frac{2\sin^2\theta_W\,s\,m_Z^2}{4m_W^2(s-m_Z^2)}\,.
\end{align}
These are proportional to the $Z$ couplings of left- and right-handed electrons, $-2g_L^e=1-2\sin^2\theta_W$ and $2g_R^e=2\sin^2\theta_W$, and to $m_Z^2/(s-m_Z^2)=s/(s-m_Z^2)-1$, the relative difference between the $Z$ and photon propagators. Integrating the $t$-channel neutrino propagator, which is proportional to $(1+\beta^2-2\beta\cos\Theta)^{-1}$, over the production angle produces the logarithm $\ln[(1+\beta)/(1-\beta)]$, which enters Eqs.~\eqref{eq:eeWWD}--\eqref{eq:eeWWA} through the function
\begin{align}\label{eq:eeWWlog}
J(x)&=\frac{1}{x}\left[
\frac{1}{\sqrt{x}}\ln\frac{1+\sqrt{x}}{1-\sqrt{x}}-2\right].
\end{align}
Up to a common factor that cancels in the fractions, the angle-integrated total cross section $\hat\sigma_{\rm tot}$, longitudinal cross section $\hat\sigma_{00}$, and difference $\Delta\hat\sigma=\hat\sigma_{--}-\hat\sigma_{++}$ of the two transverse cross sections of the $W^-$ are
\begin{align}\label{eq:eeWWD}
\hat\sigma_{\rm tot}={}&\frac{2x}{3}(3x^2-26x+27)(a^2+b^2)\nonumber\\
&-\frac{2x}{3}(3x^2-30x+31)a\nonumber\\
&+2x^2-12x+16\nonumber\\
&+x\big[x^2-6x+13+(1-x)^2(9-x)a\big]J(x),
\\[2pt]\label{eq:eeWWN}
\hat\sigma_{00}={}&\frac{2x}{3}(x^2-14x+17)(a^2+b^2)\nonumber\\
&-\frac{1}{3}(3x^3-35x^2+41x-1)a\nonumber\\
&+x^2-\frac{22x}{3}+7\nonumber\\
&-\frac{(1-x)^2}{2}\big[(x^2-10x+1)a+5-x\big]J(x),
\\[2pt]\label{eq:eeWWA}
\Delta\hat\sigma={}&\sqrt{x}\,\Big\{
2\big[a(x^2-1)+3-x\big]\nonumber\\
&+\big[(1-x)^2(x+3)a+7+2x-x^2\big]J(x)\Big\}.
\end{align}
The angle-integrated fractions then read
\begin{align}\label{eq:eeWWclosed}
\rho_{00}^{W^-}&=\frac{\hat\sigma_{00}}{\hat\sigma_{\rm tot}},\nonumber\\
\rho_{--}^{W^-}&=\frac{\hat\sigma_{\rm tot}-\hat\sigma_{00}+\Delta\hat\sigma}{2\hat\sigma_{\rm tot}},
&\rho_{++}^{W^-}&=\frac{\hat\sigma_{\rm tot}-\hat\sigma_{00}-\Delta\hat\sigma}{2\hat\sigma_{\rm tot}}.
\end{align}
For $W^+$, CP symmetry gives
\begin{equation}\label{eq:eeWWcharge}
\rho_{00}^{W^+}=\rho_{00}^{W^-},\qquad
\rho_{\pm\pm}^{W^+}=\rho_{\mp\mp}^{W^-}.
\end{equation}
For numerical evaluation near threshold, $J(x)=2/3+2x/5+\mathcal O(x^2)$; all three fractions tend to $1/3$ as $s\to4m_W^2$ from above. The helicity axis is undefined exactly at threshold. The limit $1/3$ does not mean that the $W$ bosons are produced unpolarized. Near threshold the $t$-channel neutrino exchange dominates and produces the pair in an $S$ wave from left-handed electrons, so that the spin projection of the $W^-$ on the $e^-$ beam axis is $0$ or $-1$, each with probability $1/2$, and never $+1$. In an $S$ wave, however, the $W$ direction, which defines the helicity axis, is uncorrelated with the $W$ spin, and integrating over the production angle averages the helicity projections over all directions, which gives $1/3$ for each helicity. For observables that depend only on these fractions, such as the boosted EEC of Eq.~\eqref{eq:rhocontractionboost}, the $W$ bosons near threshold are therefore indistinguishable from unpolarized ones, although their spin remains polarized along the beam axis.

Tab.~\ref{tab:WWfractions} uses the $G_\mu$-scheme inputs of Sec.~\ref{sec:ppprod}, $m_W=80.385$~GeV, $m_Z=91.1876$~GeV, and $\sin^2\theta_W=1-m_W^2/m_Z^2\simeq0.2229$. The fractions depend on the electroweak couplings only through $\sin^2\theta_W$, since the overall coupling cancels in the ratios.

\section{One-point correlators of polarized bosons and their boost}
\label{app:onepoint}

The simplest correlators of a polarized boson involve a single detector. 
They contain only the transverse spin $m=0$ operators, their structure functions are numbers rather than functions, and their boost requires no convolution. Precision studies of the single energy detector correlator~\cite{Riembau:2025wjc} and of the single charge detector~\cite{Lee:2026bcq} exist, both in the rest frame of the decaying boson. If we study the orientation of the single detector in the hadronic decay of a boosted boson relative to the boost direction and to the production plane, we become sensitive to the diagonal and the off-diagonal entries of the density matrix. The one-point energy correlator of a boosted electroweak boson was first considered in Ref.~\cite{Ricci:2022htc}, which showed that it discriminates between longitudinal and transverse modes and furthermore gives sensitivity to the interference between different polarizations.

We consider the energy flow $\Ecal(\vec n)$ of Eq.~\eqref{eq:Edef} and a charge flux
\begin{align}\label{eq:fluxdef}
\Fcal(\vec n)=\sum_{i}Q_i\,\delta^2(\hat p_i-\vec n)\,,
\end{align}
which counts the particles in the direction $\vec n$ weighted by a charge $Q_i$, for example the electric charge or a flavor quantum number such as the heavy-flavor number~\cite{Lee:2026bcq}.
In the rest frame we define, in analogy with Eq.~\eqref{eq:wdef},
\begin{align}\label{eq:onepointdef}
\Wcal^{\Ecal}_{\lambda\lambda'}(\vec n)&\equiv\frac{4\pi}{m_V}\langle V,\lambda|\Ecal(\vec n)|V,\lambda'\rangle\,,\nonumber\\
\Wcal^{\Fcal}_{\lambda\lambda'}(\vec n)&\equiv4\pi\,\langle V,\lambda|\Fcal(\vec n)|V,\lambda'\rangle\,,
\end{align}
so that the diagonal entries of $\Wcal^\Ecal$ integrate to one over $\dd\Omega/4\pi$ by the energy sum rule.
With a single direction, rotations about $\vec n$ leave the configuration invariant, so only the $m=0$ operators $\mathbf 1$, $\vec n\cdot\vec J$ and $3(\vec n\cdot\vec J)^2-2$ can appear, corresponding to $j=0,1,2$, respectively, and their structure functions cannot depend on any angle.

For the energy flow of either a $W$ or a $Z$, CP invariance of the decay, together with the fact that the energy flow does not distinguish quarks from antiquarks, gives $\Wcal^\Ecal(-\vec n)=\Wcal^\Ecal(\vec n)$, which removes $\vec n\cdot\vec J$.
For the charge flux we take a self-conjugate boson such as the $Z$, for which CP invariance gives $\Wcal^\Fcal(-\vec n)=-\Wcal^\Fcal(\vec n)$ and leaves only $\vec n\cdot\vec J$.
Hermiticity makes the structure functions real, and the energy sum rule fixes the coefficient of the identity to one.
For a $W$, the flux of a charge with nonzero total, such as the electric charge, acquires rank-zero and rank-two terms as well, so that the second line of Eq.~\eqref{eq:onepointforms} below refers to the $Z$.
Hence
\begin{align}\label{eq:onepointforms}
\Wcal^{\Ecal}(\vec n)&=\mathbf 1-a_{\Ecal}\,\big[3(\vec n\cdot\vec J)^2-2\big]\,,\nonumber\\
\Wcal^{\Fcal}(\vec n)&=a_{\Fcal}\,\vec n\cdot\vec J\,,\qquad\text{(for the $Z$)}
\end{align}
with the sign convention of Eq.~\eqref{eq:Wrotatedops}.
The parameter $a_\Ecal$ is the one of Ref.~\cite{Riembau:2025wjc}, and $a_\Fcal$ is the corresponding parameter of the charge flux of Ref.~\cite{Lee:2026bcq}. Their values and QCD corrections are discussed in these references. The positivity of the energy flux requires $-\tfrac12\leq a_\Ecal\leq1$~\mbox{\cite{Hofman:2008ar}}. The lower bound is saturated by free fermions and the upper bound by free scalars. On the other hand, $a_\Fcal$ is related to the mixed electroweak anomaly of the charge current~\cite{Hofman:2008ar,Zhang:2026emt,Zhang:2026vdo,Zhang:2026esd,Lee:2026bcq}.
Writing Eq.~\eqref{eq:onepointforms} in the helicity basis, with $\vartheta$ and $\Phi$ the polar angle and azimuth of $\vec n$, gives the entries
\begin{align}\label{eq:onepointentries}
\Wcal^\Ecal_{00}&=1+a_\Ecal(3\cos^2\vartheta-1)\,,\nonumber\\
\Wcal^\Ecal_{\pm\pm}&=\tfrac12\big(3-\Wcal^\Ecal_{00}\big)\,,\nonumber\\
\Wcal^\Ecal_{+-}&=-\tfrac32\,a_\Ecal\sin^2\vartheta\,e^{-2i\Phi}\,,\nonumber\\
\Wcal^\Ecal_{+0}&=-\Wcal^{\Ecal\,*}_{-0}=-\tfrac{3}{\sqrt2}\,a_\Ecal\sin\vartheta\cos\vartheta\,e^{-i\Phi}\,,\nonumber\\
\Wcal^\Fcal_{\pm\pm}&=\pm a_\Fcal\cos\vartheta\,,\qquad \Wcal^\Fcal_{00}=\Wcal^\Fcal_{+-}=0\,,\nonumber\\
\Wcal^\Fcal_{+0}&=\Wcal^{\Fcal\,*}_{-0}=\tfrac{1}{\sqrt2}\,a_\Fcal\sin\vartheta\,e^{-i\Phi}\,,
\end{align}
with the remaining entries fixed by Hermiticity. Although Eq.~\eqref{eq:onepointforms} depends on $\vec n$ only through $\vec n\cdot\vec J$, its helicity-basis entries depend on both angles of $\vec n$, since $\vec n\cdot\vec J=\cos\vartheta\,J_z+\tfrac12\sin\vartheta\,\big(e^{-i\Phi}J_++e^{i\Phi}J_-\big)$ with $J_\pm=J_x\pm iJ_y$. The diagonal entries depend on $\vartheta$ alone, while the entries with $\lambda-\lambda'=1$ and $2$ carry the phases $e^{-i\Phi}$ and $e^{-2i\Phi}$.
The energy-flow entries are Eqs.~\eqref{eq:wdiag} and~\eqref{eq:woffdiagrest} with $\Wcal^{(2,2)}\to0$ and $(\Wcal^{(0,0)},\Wcal^{(2,0)})\to(1,a_\Ecal)$, while the charge-flux entries are the rank-one ($j=1$) structure that the energy correlators never have.

For a boson produced with the density matrix $\rho$ of Eq.~\eqref{eq:rhocontraction}, the measured distributions are $\operatorname{Tr}[\rho\,\Wcal^\Ecal]$ and $\operatorname{Tr}[\rho\,\Wcal^\Fcal]$, and with Eq.~\eqref{eq:onepointforms} they require only two traces,
\begin{align}\label{eq:onepointtraces}
\operatorname{Tr}\big[\rho\,(\vec n\cdot\vec J)^2\big]&=\rho_{00}\sin^2\vartheta+(1-\rho_{00})\tfrac12(1+\cos^2\vartheta)\nonumber\\
&\quad+\sin^2\vartheta\,\operatorname{Re}\big[\rho_{+-}e^{2i\Phi}\big]\nonumber\\
&\quad+\sqrt2\sin\vartheta\cos\vartheta\,\operatorname{Re}\big[(\rho_{+0}-\rho_{0-})\,e^{i\Phi}\big]\,,\nonumber\\
\operatorname{Tr}\big[\rho\,\vec n\cdot\vec J\big]&=(\rho_{++}-\rho_{--})\cos\vartheta\nonumber\\
&\quad+\sqrt2\sin\vartheta\,\operatorname{Re}\big[(\rho_{+0}+\rho_{0-})\,e^{i\Phi}\big]\,.
\end{align}
The energy flow is parity even and probes the longitudinal fraction $\rho_{00}$, the alignment $\rho_{+-}$ through $\cos2\Phi$ and $\sin2\Phi$, and the combination $\rho_{+0}-\rho_{0-}$ through $\cos\Phi$ and $\sin\Phi$.
The charge flux is parity odd and probes the complementary entries, the vector polarization $\rho_{++}-\rho_{--}$ through $\cos\vartheta$ and the combination $\rho_{+0}+\rho_{0-}$ through $\cos\Phi$ and $\sin\Phi$.
Therefore, together they determine all eight real parameters of $\rho$, with $a_\Ecal$ and $a_\Fcal$ as the analyzing powers, in the same parity-even and parity-odd split as the angular coefficients of the leptonic decay~\cite{Mirkes:1992hu,Mirkes:1994eb}.

At LEP at the $Z$ pole, the $Z$ is at rest and the helicity axis is conventionally taken along the beam axis, which we take along the $e^-$ direction.
For unpolarized beams and massless electrons, angular momentum conservation along the beam forbids $\rho_{00}$ and the off-diagonal entries, and $\rho_{--}-\rho_{++}=A_e$ with $A_e=2g_V^eg_A^e/[(g_V^e)^2+(g_A^e)^2]$.
Eq.~\eqref{eq:onepointtraces} then gives
\begin{align}\label{eq:onepointLEP}
\operatorname{Tr}\big[\rho\,\Wcal^\Ecal\big]&=1-\tfrac12a_\Ecal(3\cos^2\vartheta-1)\,,\nonumber\\
\operatorname{Tr}\big[\rho\,\Wcal^\Fcal\big]&=-A_e\,a_\Fcal\cos\vartheta\,\qquad\text{(unpolarized beams)}\,.
\end{align}
The first is the  energy pattern of two-jet events, and the second is the forward-backward asymmetry of the charge flux.
Each distribution determines a single number, $a_\Ecal$ or $a_\Fcal$. 

We now return to a general density matrix and to the helicity axis of the main text, the direction of the boson momentum, and let the boson move along this axis with velocity $\beta$, $\gamma=(1-\beta^2)^{-1/2}$.
A rest-frame direction with polar angle $\vartheta$ maps to a laboratory direction with the same azimuth and
\begin{align}\label{eq:onepointaberration}
\cos\vartheta'&=\frac{\cos\vartheta+\beta}{1+\beta\cos\vartheta}\,,\nonumber\\
\lambda&=\gamma(1+\beta\cos\vartheta)=\frac{1}{\gamma(1-\beta\cos\vartheta')}\,,
\end{align}
with $\lambda$ the Doppler factor of Eq.~\eqref{eq:boosttransformation}.
Here $\hat z$ is the direction of the boson momentum, so that $\lambda$ is largest for detectors near the boson direction.
Since one detector has no internal angle, the boost acts pointwise.
Using Eq.~\eqref{eq:boosttransformation} for the energy detector, its analogue $U^\dagger(\Lambda)\Fcal(\vec n')U(\Lambda)=\lambda^2\Fcal(\vec n)$ for the charge detector, which carries no energy weight, and $|V,\lambda;\gamma\rangle=U(\Lambda)|V,\lambda\rangle$,
\begin{align}\label{eq:onepointboost}
\Wcal^{\Ecal,{\rm boost}}_{\lambda\lambda'}(\vec n';\gamma)&\equiv\frac{4\pi}{\gamma m_V}\langle V,\lambda;\gamma|\Ecal(\vec n')|V,\lambda';\gamma\rangle\nonumber\\
&=\frac{\lambda^3}{\gamma}\,\Wcal^\Ecal_{\lambda\lambda'}(\vec n)\,,\nonumber\\
\Wcal^{\Fcal,{\rm boost}}_{\lambda\lambda'}(\vec n';\gamma)&\equiv4\pi\,\langle V,\lambda;\gamma|\Fcal(\vec n')|V,\lambda';\gamma\rangle\nonumber\\
&=\lambda^2\,\Wcal^\Fcal_{\lambda\lambda'}(\vec n)\,,
\end{align}
where $\vec n$ is the rest-frame direction that maps to $\vec n'$, the powers of $\lambda$ count one energy weight and the solid-angle Jacobian, and the helicity indices are untouched because the boost preserves helicity.
The normalization is again $\int\dd\Omega'/4\pi\,\Wcal^{\Ecal,\rm boost}_{\lambda\lambda}=1$, since $\lambda^3\dd\Omega'=\lambda\,\dd\Omega$ and the rest-frame energy flow carries no net momentum, and the total charge flux is conserved.
The phases $e^{\pm i\Phi}$ and $e^{\pm2i\Phi}$ survive the boost unchanged because it commutes with rotations about the axis, so a boosted boson with an off-diagonal $\rho$ shows the same azimuthal modulations as at rest, compressed in polar angle.
In particular, the $\cos\Phi$, $\sin\Phi$, $\cos2\Phi$, and $\sin2\Phi$ modulations of Eq.~\eqref{eq:onepointtraces} about the boson direction, which measure the off-diagonal entries of $\rho$, persist in the boosted jet with the same coefficients $\rho_{+-}$ and $\rho_{+0}\mp\rho_{0-}$, so that the orientation of a single detector relative to the production plane measures them directly.
Explicitly, applying Eq.~\eqref{eq:onepointboost} to Eq.~\eqref{eq:onepointforms} with $\cos\vartheta=(\cos\vartheta'-\beta)/(1-\beta\cos\vartheta')$ and $\sin\vartheta=\sin\vartheta'/[\gamma(1-\beta\cos\vartheta')]$ inserted in $\vec n\cdot\vec J$ gives
\begin{align}\label{eq:onepointboostexplicit}
\Wcal^{\Ecal,{\rm boost}}&=\frac{\mathbf 1-a_\Ecal\big[3(\vec n\cdot\vec J)^2-2\big]}{\gamma^4(1-\beta\cos\vartheta')^3}\,,\nonumber\\
\Wcal^{\Fcal,{\rm boost}}&=\frac{a_\Fcal\,\vec n\cdot\vec J}{\gamma^2(1-\beta\cos\vartheta')^2}\,.
\end{align}
Together they still determine all eight real parameters of the density matrix $\rho$ of the boosted boson.
For a diagonal $\rho$, the boosted distributions $\operatorname{Tr}[\rho\,\Wcal^{\Ecal,\rm boost}]$ and $\operatorname{Tr}[\rho\,\Wcal^{\Fcal,\rm boost}]$ are independent of $\Phi$.

\bibliography{refs}
 
\end{document}